\documentclass[11pt, letterpaper]{article}

\usepackage[latin1]{inputenc}
\usepackage{graphicx}
\usepackage{amsfonts,amsmath,amsopn,amssymb,amsthm,bbm,latexsym,mathrsfs,pstricks,verbatim,caption,pst-node,stmaryrd}
\newcommand{\gs}{\text{gs}}

\newcommand{\group}[1]{\left(#1\right)}

\newcommand{\qbin}[2]{\left[\begin{array}{c}#1\\#2\end{array}\right]}
\let\rw\rightarrow

\let\t\

\newcommand{\su}[1]{\widehat{su}(2)_{#1}}

\let\ga\gamma
\def\M{{\cal M}}

\newcommand{\mk}{\mathcal{S}}
\newcommand{\mkp}[1]{\mathcal{S}^{(#1)}}

\newcommand{\ym}{\gamma}
\newcommand{\ymp}[1]{\ga_{#1}}

\newcommand{\wm}{\hat{w}}
\newcommand{\wmp}[1]{\hat{w}_{#1}}

\newcommand{\cm}{s}

\newcommand{\wt}{\tilde{w}}

\newcommand{\akp}[1]{\mathcal{P}^{(#1)}}

\newcommand{\wa}{w}
\newcommand{\wap}[1]{w_{#1}}

\newcommand{\ca}{s}

\newcommand{\ab}{b}
\newcommand{\abp}[2]{b_{#1}^{#2}}

\newcommand{\cip}[1]{c_{#1}}

\newcommand{\ckp}[1]{\mathcal{J}^{(#1)}}

\newcommand{\wc}{\tilde w}
\newcommand{\wcp}[1]{\tilde w_{#1}}

\let\n\noindent

\let\l\left
\let\r\right
\let\y\infty
\let\e\epsilon
\let\lrw\leftrightarrow

\newcommand{\jv}{\vec{\jmath}}
\newcommand{\jh}{\hat{\jmath}}
\let\n\noindent

\title{\vskip60pt {Path representation of $\su{k}$ states II:}
\vskip1pt{Operator construction of the fermionic character and spin-$\frac12$--RSOS factorization}}

\author{\bf{Jo\"el Lamy-Poirier\footnote{Present address: Perimeter Institute for Theoretical Physics, Waterloo, Ontario, N2J 2Y5, Canada.} 
and  Pierre
Mathieu} \\ 
\\
D\'epartement de physique, de g\'enie physique et d'optique,\\
Universit\'e Laval,
Qu\'ebec, Canada, G1V 0A6.\\
 jlamypoirier@perimeterinstitute.ca,
pmathieu@phy.ulaval.ca}

\begin{document}
\maketitle

\begin{abstract}
This is the second of two articles (independent of each other) devoted to the analysis of the path description of the states in $\su{k}$ WZW models. Here we present a constructive derivation of the fermionic character at level $k$  based on these paths. 
The starting point  is the expression of a path in terms of a sequence of nonlocal (formal) operators acting on the vacuum ground-state path. 
Within this framework, the key step  is the construction of the level-$k$ operator sequences out of those at  level-1 by the action of a new type of operators. These actions of operators on operators turn out to have a path interpretation: these paths are precisely the finitized RSOS paths related to the unitary minimal models $\mathcal{ M}(k+1,k+2)$. 
We thus unravel -- at the level of the path representation of the states --, a direct factorization into a $k=1$ spinon part times a RSOS factor.
It is also  pointed out that since there are two fermionic forms describing these finite RSOS paths, the resulting fermionic $\su{k}$ characters arise in two versions. Finally, the relation between the present construction and the Nagoya spectral decomposition of the path space is sketched.\end{abstract} 

\newpage
\section{Introduction}

\subsection{Motivation}

The fermionic characters in conformal field theory \cite{KKMM} reflects an underlying quasi-particle formulation that should capture some information concerning the massive particles characterizing a given integrable perturbation \cite{Zam}. 
An indication for a close relation between a given fermionic form and a particular integrable perturbation is rooted in the practical observation that the most efficient way of obtaining these fermionic forms is actually to start with a path description of the states  inherited from  given off-critical integrable statistical-model version of the conformal field theory under consideration.\footnote{This observation relies on the highly nontrivial fact that the generating functions for the one-dimensional configurations, configurations that define the paths (and we stress that these configuration sums which are defined off-criticality), are characters of the conformal theory. This  was first observed in \cite{Kyoto} and is now understood in terms of the quantum-group symmetry of the integrable lattice models: the state counting is $q$-independent ($q=1$ corresponding to the critical point).} 
For instance, the complete fermionic description of the minimal models obtained in \cite{Wel}  is based on the paths associated to the configuration of the RSOS models in regime III \cite{ABF,FB}, the lattice representation  of their
$\phi_{1,3}$-perturbation. 

Quite interestingly, the paths can generally be decomposed into some basic building blocks, often called  (path-)particles \cite{BreL,OleJS,PMjmp}. In some cases, these seem to correspond to the massless versions of the off-critical massive particles. For instance,
the particle content of the restricted sine-Gordon model which, for a suitable value of the coupling constant, is the scaling limit of a $\phi_{1,3}$-perturbed minimal model  
\cite{LeC,Smi}, can be read directly from the particle decomposition of the (regime III) RSOS paths
\cite{JMpar}. A more spectacular example is the recent decomposition of the paths of the dilute $A_3$ model \cite{dil}, the lattice representation of the $\phi_{1,2}$ perturbed Ising model, in terms of eight particles whose basic properties are controlled by the $E_8$ Cartan matrix \cite{OW}, in conformity with the $E_8$ particle-spectrum derived by Zamolodchikov from $S$ matrices \cite{Zam8}.

However, some reservations are in order: the path-particle decomposition is generically not unique. The analysis of the $\su{2}$ WZW model in \cite{LPM} illustrates this point very neatly: the paths can be described in terms of three or four particles. Moreover, still another particle description is furnished here. Therefore, if this relationship can be made reliable, there is a missing criterion selecting the proper physical decomposition, the clue for which is best investigated by the consideration of more examples.\footnote{The non-uniqueness of fermionic forms associated to a given perturbation has been observed long ago; see for instance \cite{BMP}.}

Here we are interested in the $\su{k}$ WZW model (see e.g., \cite{CFT}).
The spectrum of the related XXX chain for spin higher than $\frac12$ has first been obtained via a Bethe-ansatz analysis of the scattering matrices \cite{Res}. These scattering matrices show a remarkable factorization pattern into a spin-$\frac12$ piece times a RSOS factor. The same underlying particle structure emerges from an analysis of the six-vertex model considered from the point of view of the affine quantum group symmetry \cite{Idz}. 
Independently, the tensor-product form of the scattering matrices for the (current-current)  perturbed $\su{k}$ WZW  model  have been derived from general symmetry arguments in \cite{ABL} (see also \cite{Fen}).

These gross features show up in the WZW fermionic characters. These characters were first obtained from a conjectured basis of states in \cite{BLSb}, and their structure indeed reflects such a spin-$\frac12$--RSOS factorization pattern.
These character formulae were subsequently proved in
\cite{Ara,NYa}.

The original motivation for the present work was to re-derive the fermionic $\su{k}$ characters using the representation of the states in terms of paths \cite{DJKMO3,JMMO}, in the spirit of the constructive method of \cite{OleJS,BreL}, with the eventual goal of unravelling the path-particle structure.
 Our approach led us directly to a description of a path at level $k$ in terms of two components: a level-1 path and a RSOS path, both of which have a clear particle structure (cf. \cite{LPM} and \cite{OleJS} respectively).\footnote{The relationship with the construction in \cite{Ara} is briefly discussed at the end of the following subsection and a more detailed analysis is presented in App. \ref{Nag}.}

\subsection{Organization of the work}
The  paths describing the $\su{k}$ states, whose set is denoted by $\mkp{k}$, are defined in Section \ref{sk def}  \cite{DJKMO3,JMMO,Ara}. Our sole input at this stage is to introduce for them a new weight function that greatly simplifies their analysis. Its demonstration is given at the end of Section \ref{carop}.
   
 The starting point of our analysis is the plain observation that any path can be described by the action of a sequence of nonlocal operators $b^{\pm1}$  (which are formal operators that maps a path to another path) on the level-$k$ vacuum ground-state path (cf. Section \ref{bk}). This construction differs from the $k=1$ version introduced in \cite{LPM} in that an operator $b^{\pm1}$ can act repeatedly, but $\leq  k$ times, at the same position. Constructing the characters amount to $q$-enumerate such sequences.

A convenient way to attack this problem is to consider the level-$k$ sequences as built up from a class of restricted ones, the latter being in 1-1 correspondence with paths at level 1   (cf. Section \ref{Strategy}).
  These restricted sequences are given by strings of operators  $b_l^{a}$ (where the subindex indicates the action position) such that $|a|\leq 1$; in addition, precise distance conditions must be satisfied between the application point of  two successive operators. Now, in order to generate level-$k$ sequences out of the restricted ones, 
we introduce still another type of operators, denoted by $c$,  acting directly on sequences of $b$-type operators. 
Roughly, these operators $c$ modify the application position of a string of nonlocal operators with the result that not only the distance conditions are violated
but the number of applications of an operator at a given position can increase (for instance it might transform  $b_{l-1}^{a}b_l$ into  $b_{l-1}^{a+1}$).

The next step is the observation that the action of $c$-type operators on restricted sequences can be  interpreted in terms of paths (cf. Section \ref{chejk}). 
 These paths are actually finite RSOS paths, 
 representatives of the (finitized) states of the $\M(k+1,k+2)$ minimal models (and recall that $\M(2,3)$ is trivial). 

 Once this correspondence is established, 
 one can use the results of    \cite{OleJS,FW} to write down the generating function of these finite RSOS paths. Actually, this generating function comes in two versions \cite{Mel2}. These results are presented in Section \ref{generc}. The full generating function takes the form of  a product of a $k=1$ piece, capturing the $q$-enumeration of the restricted sequences with fixed operator content, times a factor enumerating RSOS paths with length equal to the total number of operators, and the product is summed over the number of operators of each type. The fermionic character of \cite{BLSb} is then recovered in Section \ref{fermica},
 together with a new Melzer-type variant.

Let us now comment on the relation with the work \cite{Ara}. It is shown there that a collection of $\mkp{k}$ paths with the same `local energies' can be associated with a Yangian piece (represented there by a Young diagram) and a RSOS path. Since the former part can be viewed as a collection of level-1 paths organized in  a number of $su(2)$ multiplets, a relation with the present construction is quite likely. Indeed, we show in App. \ref{Nag} that the  RSOS path associated to a given $\ga$ path is the same in the two constructions.

Finally, we should stress that the present work can be read independently of its first part  \cite{LPM}, devoted to the detailed study of two special cases: $k=1,2$. 

\section{Path description of $\su{k}$ states}\label{sk def}

$\mkp{k}$ paths are specified by the sequence of integers $(\ga_0,\ga_1,\ga_2,\cdots)$ subject to the restriction:\footnote{Compared to  \cite{DJKMO3,JMMO,Ara} and  \cite{LPM} for $k=1,2$, we have introduced in the definition of $\mkp{k}$ paths an extra vertex at $l=0$ with a prescribed value, $\ga_0=-k$. This addition of this pre-segment was motivated in  \cite[Section 2.4.3]{LPM} by its naturalness in the particle interpretation of the paths. Here, it is justified by the relation (\ref{gavsa}) below that links a path to its operator description, whose validity can thereby be extended to the case $l=0$.}
\begin{align}
	\ymp{l} \in\{-k,-k+2,\cdots, k-2,k\} \qquad\text{and}\qquad \ga_0=-k.
\end{align}
The paths are subdivided into classes $\mkp{k} _j$ depending on their values of $j$, 
with $0\leq j\leq k/2$. 
The elements of $\mkp{k}_j$ are such that they eventually oscillate between the values $k-4j$ and $-(k-4j)$, matching the tail of the ground-state path  $\ym^{\gs(j;k)}\in \mkp{k}_j$ defined as

\footnote{This state is $(2j+1)$-degenerated with respect to the weight but this particular one is marked out by its sector value equal to $-j$ (see below), and its purely oscillatory behavior. The different paths of the multiplet are identical except for their value of $\ga_1$.}
\begin{align}
	\ymp{l}^{\gs(j;k)} = (-1)^{l-1}(k-4j).\label{groundk}
\end{align}
Note that this path is a straight line  on the axis if $k$ is even and $j=k/4$ (see Fig. \ref{exS4} below).

It is convenient to introduce the auxiliary values $j_l$ defined as follows:
\begin{align}
	j_l   	=\frac{1}{4}(k+(-1)^l\ymp{l}) 
	=\frac{1}{4}|\ymp{l}-(-1)^{l-1}k|
	=\frac{1}{4}|\ymp{l}-\ym_{l}^{\gs(0;k)}|
	\label{defdejl},
	\end{align}
where we used $\ymp{l}\leq k$ to obtain the second equality. The set of the values $j_l$ for a given path will be denoted compactly by the vector $\jv\equiv (j_0,j_1,j_2,\cdots)$.	The class parameter $j$ is recovered from the limiting value
	\begin{align}
	 j = \lim_{l\rw\infty}j_l.\label{defdej}
\end{align}

The sector of a path $\ga\in\mkp{k}_j$ is given by:
\begin{align}
	\cm &= -j + \frac{1}{2}\sum_{l=1}^{\infty}(\ymp{l}-\ym_{l}^{\gs(j;k)}),\label{cmlk}
\end{align}
from which it follows that $s(\ga^{\gs(j;k)})=-j$. Moreover, this also implies that $s+j\in\mathbb Z$.

Finally, the weight of a  path $\ga\in\mkp{k}_j$ is given by 
\begin{align}\label{weightmk}
	\wm(\ga)&=\sum_{l=1}^{\infty}(\wmp{l}^{\circ}(\ga)-\wmp{l}^{\circ}(\ga^{\gs(j;k)})),
	\nonumber\\
	\wmp{l}^{\circ}(\ga)&=\frac12l[k-\min(\ymp{l},-\ymp{l+1})].
\end{align}
This expression for the weight is essentially equivalent to the one given in \cite{DJKMO3,JMMO,Ara}.
In the following analysis, we  mainly use the following alternative expression for the weight function for a $\ga$ path:\footnote{The summations in (\ref{cmlk}), (\ref{weightmk}) and (\ref{weightak}) formally start at $l=0$ but in the three cases, the $l=0$ term does not contribute.}
\begin{align}\label{weightak}
	\wa(\ga)&=-\frac12(\ca+j)+\sum_{l=1}^{\infty}\wap{l}(\ga),\nonumber\\
	\wap{l}(\ga)&=\frac{l}4|\ymp{l}+\ymp{l+1}|=l|j_{l+1}-j_l|.
\end{align}
We postpone the proof  of the equivalence of the weights (\ref{weightmk}) and (\ref{weightak})
to the end of Section \ref{carop}.

The  result of \cite{DJKMO3} can be phrased as follows: the character $\chi_{j;k}(z;q)$ of the level-$k$ integrable module with finite Dynkin label $2j$ is
\begin{equation}\label{cagf}
\chi_{j;k}(z;q)=\sum_{\ga\in\mkp{k}_j}q^{w(\ga)}\, z^{s(\ga)}.
\end{equation}

\begin{figure}[ht]\
\caption{Example of a $\mk^{(3)}_0$ path, with $\cm=-1$ and $\wa=21$. 
For $l\geq8$, the path matches $\ym^{\gs(0;3)}$, drawn in dotted lines.}
\label{exemplemk}
\begin{center}
\begin{pspicture}(3,0)(8,3)
{\psset{yunit=20pt,xunit=20pt,linewidth=.8pt}
\psline{->}(0,0)(0,3.5) \psline{->}(0,1.5)(15,1.5)
\rput(2,1){{\scriptsize $2$}}
\rput(1,1){{\scriptsize $1$}}
\rput(3,1){{\scriptsize $3$}}
\rput(7,1){{\scriptsize $7$}}
\rput(9,1){{\scriptsize $9$}}
\rput(11,1){{\scriptsize $11$}}
\rput(13,1){{\scriptsize $13$}}
\rput(4,1){{\scriptsize $4$}}
\rput(6,1){{\scriptsize $6$}}
\rput(8,1){{\scriptsize $8$}}
\rput(10,1){{\scriptsize $10$}}
\rput(12,1){{\scriptsize $12$}}
\rput(14,1){{\scriptsize $14$}}

\rput[r](-0.2,0){{\scriptsize $-3$}}
\rput[r](-0.2,1){{\scriptsize $-1$}}
\rput[r](-0.2,2){{\scriptsize $1$}}
\rput[r](-0.2,3){{\scriptsize $3$}}

\psset{linestyle=solid}
\psline{-}(1,1.4)(1,1.6)\psline{-}(2,1.4)(2,1.6)
\psline{-}(3,1.4)(3,1.6)\psline{-}(4,1.4)(4,1.6)
\psline{-}(5,1.4)(5,1.6)\psline{-}(6,1.4)(6,1.6)
\psline{-}(7,1.4)(7,1.6)\psline{-}(8,1.4)(8,1.6)
\psline{-}(9,1.4)(9,1.6)\psline{-}(10,1.4)(10,1.6)
\psline{-}(11,1.4)(11,1.6)\psline{-}(12,1.4)(12,1.6)
\psline{-}(13,1.4)(13,1.6)\psline{-}(14,1.4)(14,1.6)

\psline{-}(0,1)(.1,1)
\psline{-}(0,2)(.1,2)
\psline{-}(0,3)(.1,3)

\psline{-}(0,0)(1,0)
\psline{-}(1,0)(2,3)
\psline{-}(2,3)(3,2)\psline{-}(3,2)(4,3)
\psline{-}(4,3)(5,1)\psline{-}(5,1)(6,2)
\psline{-}(6,2)(7,0)\psline{-}(7,0)(8,0)
\psline{-}(8,0)(9,3)\psline{-}(9,3)(10,0)
\psline{-}(10,0)(11,3)\psline{-}(11,3)(12,0)
\psline{-}(12,0)(13,3)\psline{-}(13,3)(14,0)

\psset{linestyle=dashed,dashadjust=false}
\psline{-}(14,0)(15,3)

\psset{linestyle=dotted}
\psline{-}(0,0)(1,3)
\psline{-}(1,3)(2,0)
\psline{-}(2,0)(3,3)
\psline{-}(3,3)(4,0)
\psline{-}(4,0)(5,3)
\psline{-}(5,3)(6,0)
\psline{-}(6,0)(7,3)
\psline{-}(7,3)(8,0)

}
\end{pspicture}
\end{center}
\end{figure}
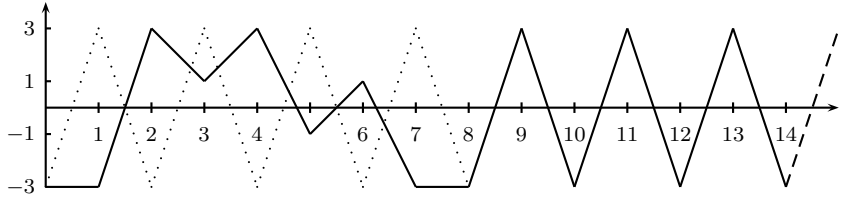

We illustrate these paths with an example $\ym\in\mkp{3}$ represented on Fig. \ref{exemplemk}, that is
\begin{align}
	\ym =(-3,-3,3,1,3,-1,1,-3,-3,3,-3,3,-3,\ldots),\label{exemplemky}
\end{align}
whose tail is that  of $\ym^{\gs(0;3)}$, which fixes $j=0$. 
The values of $j_l$, obtained by (\ref{defdej}), are: 
\begin{align}
	\jv=\group{0,\tfrac{3}{2},\tfrac{3}{2},\tfrac{1}{2},\tfrac{3}{2},1,1,\tfrac{3}{2},0,0,0,...}\qquad 
		(\Rightarrow \;j=\lim_{l\rw\infty}j_l=0).
\end{align}
For the sector, we have, using (\ref{cmlk}) with $j=0$:
\begin{align}
	\left\{\tfrac{1}{2}(\ymp{l}-\ym_{l}^{\gs(0;3)}) |~l\geq1\right \}&=(-3,3,-1,3,-2,2,-3,0,0,0,...)\quad \implies \quad s=-1.
\end{align}
Finally, the weight, calculated from (\ref{weightak}), is:
\begin{align}
               & 	\l\{\wap{l}=\tfrac{1}{4}l|\ymp{l}+\ymp{l+1}| |~l\geq1\r \}=(0,2,3,2,0,3,\tfrac{21}2,0,0,0,...),\nonumber\\
	\qquad&\implies \quad\wa=-\tfrac{1}{2}(s+j)+\sum_{l=1}^\y \wap{l}=\tfrac{1}{2}+\tfrac{41}{2}=21.
\end{align}

\section{Nonlocal operator description of $\mkp{k}$ paths}\label{bk}

We now introduce a description of the $\mkp{k}$ paths in terms of a sequence of nonlocal operators $\ab^{a}$, with $|a|\leq k$, acting  on the ground state $\ga^{\gs(0;k)}$. 
The aim of this section is to define these nonlocal operators, characterize a path by a sequence of operators and extract the path characteristics in terms of the sequence data.


\subsection{Operators on $\mkp{k}$ paths}\label{bksect}

Any path  $\ga=(-k,\ga_1,\ga_2,\cdots)\in\mkp{k}$ can be written as a sequence of operators acting on the path $\ga^{\gs(0;k)}$ successively as follows:
\begin{equation}\label{pirev}
\ga= \cdots b_{l}^{a_{l}}\cdots b_{2}^{a_{2}} b_{1}^{a_{1}} b_{0}^{a_{0}} \ga^{\gs(0;k)},\qquad \text{with}\quad \ga^{\gs(0;k)}=(-k,k,-k,k,-k,\cdots),\end{equation}
where the $a_l$ are integers bounded by $|a_l|\leq k$, with only finitely many $a_i\ne0$.
The action of $b_l^n$  on a  generic path $\ga$, with $b_l^n\ga=\ga'$, is: 
\begin{align}\label{bkdef}
	\abp{l}{n}&(\ga_0,\ymp{1}, \ymp{2}, ... , \ymp{l}, \ymp{l+1}, \ymp{l+2}, ...)=
		(\ga_0,\ymp{1}, \ymp{2}, ... , \ymp{l}, \ymp{l+1}', \ymp{l+2}', \ymp{l+3}', ...) \nonumber\\
	&	\quad \;\,\mbox{if} \;\; |\ymp{l+m}'|\leq k\;\;\forall \; m\geq1,\quad\text{where}			
				\quad\ymp{l+m}'=\ymp{l+m}-2n(-1)^{m},
\end{align}
and it is 0 otherwise. Note that $\ga_0$ is never affected by this action.

For the path $\ga$ represented by (\ref{pirev}), using $\ga_l^{\gs(0;k)}=(-1)^{l-1}k$ (cf. (\ref{groundk})), the action (\ref{bkdef}) implies that
\begin{align}\label{gal=}\ga_l=(-1)^{l-1}(k+2a_0-2a_1+2a_2+\cdots +(-1)^{l-1}2a_{l-1}).
\end{align}
This leads to  the useful relation 
\begin{equation}\label{gavsa}
\ga_{l+1}+\ga_l=2a_l.
\end{equation}
This expression (\ref{gal=}) also makes the non-vanishing conditions $|\ga_l|\leq k$ readily formulated in terms of the $a_i$ as
\begin{equation}\label{cona}
-k\leq \sum_{i=0}^m(-1)^{i} a_i\leq 0,\qquad \forall m\geq 0.
\end{equation}
Note that for $m=0$, this forces $a_0\leq 0$. In particular, the different ground states $\ym^{\gs(j;k)}$, for $j=0,\frac12,\cdots,\frac{k}2$, are  obtained as:
\begin{align}\label{gsmul}
	\ym^{\gs(j;k)}=\abp{0}{-2j}\ym^{\gs(0;k)}.
\end{align}

\begin{figure}[htpb]\
\caption{The step by step construction of the $\mkp{3}$ path displayed in Fig. \ref{exemplemk} by a sequence of $\abp{l}{a_l}$ operators acting on $\ym^{\gs(0;3)}$. This illustrates the action (\ref{bkdef}). Besides each path, we indicate the value of $j$ as computed from (\ref{jb}).}

\label{fig3}
\begin{center}
\begin{pspicture}(0,0)(3,17)
{\psset{yunit=14pt,xunit=16pt,linewidth=.8pt}
\rput[Br](-1,31){{ $\ym^{\gs(0;3)}~(j=0)$}}
\rput[Br](-1,26){{ $\abp{0}{-3}\,\ym^{\gs(0;3)}~(j=\tfrac32)$}}
\rput[Br](-1,21){{ $\abp{2}{2}\,\abp{0}{-3}\,\ym^{\gs(0;3)}~(j=\tfrac12)$}}
\rput[Br](-1,16){{ $\abp{3}{2}\,\abp{2}{2}\,\abp{0}{-3}\,\ym^{\gs(0;3)}~(j=\tfrac32)$}}
\rput[Br](-1,11){{ $\abp{4}{1}\,\abp{3}{2}\,\abp{2}{2}\,\abp{0}{-3}\,\ym^{\gs(0;3)}~(j=1)$}}
\rput[Br](-1,6){{ $\abp{6}{-1}\,\abp{4}{1}\,\abp{3}{2}\,\abp{2}{2}\,\abp{0}{-3}\,\ym^{\gs(0;3)}~(j=\tfrac32)$}}
\rput[Br](-1,1){{ $\abp{7}{-3}\,\abp{6}{-1}\,\abp{4}{1}\,\abp{3}{2}\,\abp{2}{2}\,\abp{0}{-3}\,\ym^{\gs(0;3)}~(j=0)$}}

\rput(1,1){{\scriptsize $1$}}
\rput(2,1){{\scriptsize $2$}}
\rput(3,1){{\scriptsize $3$}}
\rput(7,1){{\scriptsize $7$}}
\rput(9,1){{\scriptsize $9$}}
\rput(11,1){{\scriptsize $11$}}
\rput(13,1){{\scriptsize $13$}}
\rput(4,1){{\scriptsize $4$}}
\rput(6,1){{\scriptsize $6$}}
\rput(8,1){{\scriptsize $8$}}
\rput(10,1){{\scriptsize $10$}}
\rput(12,1){{\scriptsize $12$}}
\rput(14,1){{\scriptsize $14$}}
\psline{->}(0,-1)(0,4)

\psline{->}(0,4.2)(0,9)
\psline{->}(0,9.2)(0,14)
\psline{->}(0,14.2)(0,19)
\psline{->}(0,19.2)(0,24)
\psline{->}(0,24.2)(0,29)
\psline{->}(0,29.2)(0,34)
\psline{->}(0,1.5)(15,1.5)
\psline{->}(0,6.5)(15,6.5)
\psline{->}(0,11.5)(15,11.5)
\psline{->}(0,16.5)(15,16.5)
\psline{->}(0,21.5)(15,21.5)
\psline{->}(0,26.5)(15,26.5)
\psline{->}(0,31.5)(15,31.5)

\psline{-}(0,0)(0.2,0)\psline{-}(0,1)(0.2,1)\psline{-}(0,2)(0.2,2)\psline{-}(0,3)(0.2,3)
\psline{-}(0,5)(0.2,5)\psline{-}(0,6)(0.2,6)\psline{-}(0,7)(0.2,7)\psline{-}(0,8)(0.2,8)
\psline{-}(0,10)(0.2,10)\psline{-}(0,11)(0.2,11)\psline{-}(0,12)(0.2,12)\psline{-}(0,13)(0.2,13)
\psline{-}(0,15)(0.2,15)\psline{-}(0,16)(0.2,16)\psline{-}(0,17)(0.2,17)\psline{-}(0,18)(0.2,18)
\psline{-}(0,20)(0.2,20)\psline{-}(0,21)(0.2,21)\psline{-}(0,22)(0.2,22)\psline{-}(0,23)(0.2,23)
\psline{-}(0,25)(0.2,25)\psline{-}(0,26)(0.2,26)\psline{-}(0,27)(0.2,27)\psline{-}(0,28)(0.2,28)
\psline{-}(0,30)(0.2,30)\psline{-}(0,31)(0.2,31)\psline{-}(0,32)(0.2,32)\psline{-}(0,33)(0.2,33)

\psline{-}(1,1.4)(1,1.6)\psline{-}(2,1.4)(2,1.6)\psline{-}(3,1.4)(3,1.6)\psline{-}(4,1.4)(4,1.6)\psline{-}(5,1.4)(5,1.6)\psline{-}(6,1.4)(6,1.6)
\psline{-}(7,1.4)(7,1.6)\psline{-}(8,1.4)(8,1.6)\psline{-}(9,1.4)(9,1.6)\psline{-}(10,1.4)(10,1.6)\psline{-}(11,1.4)(11,1.6)
\psline{-}(12,1.4)(12,1.6)\psline{-}(13,1.4)(13,1.6)\psline{-}(14,1.4)(14,1.6)

\psline{-}(1,6.4)(1,6.6)\psline{-}(2,6.4)(2,6.6)\psline{-}(3,6.4)(3,6.6)\psline{-}(4,6.4)(4,6.6)\psline{-}(5,6.4)(5,6.6)\psline{-}(6,6.4)(6,6.6)
\psline{-}(7,6.4)(7,6.6)\psline{-}(8,6.4)(8,6.6)\psline{-}(9,6.4)(9,6.6)\psline{-}(10,6.4)(10,6.6)\psline{-}(11,6.4)(11,6.6)
\psline{-}(12,6.4)(12,6.6)\psline{-}(13,6.4)(13,6.6)\psline{-}(14,6.4)(14,6.6)

\psline{-}(1,11.4)(1,11.6)\psline{-}(2,11.4)(2,11.6)\psline{-}(3,11.4)(3,11.6)\psline{-}(4,11.4)(4,11.6)\psline{-}(5,11.4)(5,11.6)\psline{-}(6,11.4)(6,11.6)
\psline{-}(7,11.4)(7,11.6)\psline{-}(8,11.4)(8,11.6)\psline{-}(9,11.4)(9,11.6)\psline{-}(10,11.4)(10,11.6)\psline{-}(11,11.4)(11,11.6)
\psline{-}(12,11.4)(12,11.6)\psline{-}(13,11.4)(13,11.6)\psline{-}(14,11.4)(14,11.6)

\psline{-}(1,16.4)(1,16.6)\psline{-}(2,16.4)(2,16.6)\psline{-}(3,16.4)(3,16.6)\psline{-}(4,16.4)(4,16.6)\psline{-}(5,16.4)(5,16.6)\psline{-}(6,16.4)(6,16.6)
\psline{-}(7,16.4)(7,16.6)\psline{-}(8,16.4)(8,16.6)\psline{-}(9,16.4)(9,16.6)\psline{-}(10,16.4)(10,16.6)\psline{-}(11,16.4)(11,16.6)
\psline{-}(12,16.4)(12,16.6)\psline{-}(13,16.4)(13,16.6)\psline{-}(14,16.4)(14,16.6)

\psline{-}(1,21.4)(1,21.6)\psline{-}(2,21.4)(2,21.6)\psline{-}(3,21.4)(3,21.6)\psline{-}(4,21.4)(4,21.6)\psline{-}(5,21.4)(5,21.6)\psline{-}(6,21.4)(6,21.6)
\psline{-}(7,21.4)(7,21.6)\psline{-}(8,21.4)(8,21.6)\psline{-}(9,21.4)(9,21.6)\psline{-}(10,21.4)(10,21.6)\psline{-}(11,21.4)(11,21.6)
\psline{-}(12,21.4)(12,21.6)\psline{-}(13,21.4)(13,21.6)\psline{-}(14,21.4)(14,21.6)

\psline{-}(1,26.4)(1,26.6)\psline{-}(2,26.4)(2,26.6)\psline{-}(3,26.4)(3,26.6)\psline{-}(4,26.4)(4,26.6)\psline{-}(5,26.4)(5,26.6)\psline{-}(6,26.4)(6,26.6)
\psline{-}(7,26.4)(7,26.6)\psline{-}(8,26.4)(8,26.6)\psline{-}(9,26.4)(9,26.6)\psline{-}(10,26.4)(10,26.6)\psline{-}(11,26.4)(11,26.6)
\psline{-}(12,26.4)(12,26.6)\psline{-}(13,26.4)(13,26.6)\psline{-}(14,26.4)(14,26.6)

\psline{-}(1,31.4)(1,31.6)\psline{-}(2,31.4)(2,31.6)\psline{-}(3,31.4)(3,31.6)\psline{-}(4,31.4)(4,31.6)\psline{-}(5,31.4)(5,31.6)\psline{-}(6,31.4)(6,31.6)
\psline{-}(7,31.4)(7,31.6)\psline{-}(8,31.4)(8,31.6)\psline{-}(9,31.4)(9,31.6)\psline{-}(10,31.4)(10,31.6)\psline{-}(11,31.4)(11,31.6)
\psline{-}(12,31.4)(12,31.6)\psline{-}(13,31.4)(13,31.6)\psline{-}(14,31.4)(14,31.6)

\psline{-}(0,0)(1,0)
\psline{-}(1,0)(2,3)
\psline{-}(2,3)(3,2)
\psline{-}(3,2)(4,3)
\psline{-}(4,3)(5,1)
\psline{-}(5,1)(6,2)
\psline{-}(6,2)(7,0)
\psline{-}(7,0)(8,0)
\psline{-}(8,0)(9,3)
\psline{-}(9,3)(10,0)
\psline{-}(10,0)(11,3)
\psline{-}(11,3)(12,0)
\psline{-}(12,0)(13,3)
\psline{-}(13,3)(14,0)

\psline{-}(0,5)(1,5)
\psline{-}(1,5)(2,8)
\psline{-}(2,8)(3,7)
\psline{-}(3,7)(4,8)
\psline{-}(4,8)(5,6)
\psline{-}(5,6)(6,7)
\psline{-}(6,7)(7,5)
\psline{-}(7,5)(8,8)
\psline{-}(8,8)(9,5)
\psline{-}(9,5)(10,8)
\psline{-}(10,8)(11,5)
\psline{-}(11,5)(12,8)
\psline{-}(12,8)(13,5)
\psline{-}(13,5)(14,8)

\psline{-}(0,10)(1,10)
\psline{-}(1,10)(2,13)
\psline{-}(2,13)(3,12)
\psline{-}(3,12)(4,13)
\psline{-}(4,13)(5,11)
\psline{-}(5,11)(6,12)
\psline{-}(6,12)(7,11)
\psline{-}(7,11)(8,12)
\psline{-}(8,12)(9,11)
\psline{-}(9,11)(10,12)
\psline{-}(10,12)(11,11)
\psline{-}(11,11)(12,12)
\psline{-}(12,12)(13,11)
\psline{-}(13,11)(14,12)

\psline{-}(0,15)(1,15)
\psline{-}(1,15)(2,18)
\psline{-}(2,18)(3,17)
\psline{-}(3,17)(4,18)
\psline{-}(4,18)(5,15)
\psline{-}(5,15)(6,18)
\psline{-}(6,18)(7,15)
\psline{-}(7,15)(8,18)
\psline{-}(8,18)(9,15)
\psline{-}(9,15)(10,18)
\psline{-}(10,18)(11,15)
\psline{-}(11,15)(12,18)
\psline{-}(12,18)(13,15)
\psline{-}(13,15)(14,18)

\psline{-}(0,20)(1,20)
\psline{-}(1,20)(2,23)
\psline{-}(2,23)(3,22)
\psline{-}(3,22)(4,21)
\psline{-}(4,21)(5,22)
\psline{-}(5,22)(6,21)
\psline{-}(6,21)(7,22)
\psline{-}(7,22)(8,21)
\psline{-}(8,21)(9,22)
\psline{-}(9,22)(10,21)
\psline{-}(10,21)(11,22)
\psline{-}(11,22)(12,21)
\psline{-}(12,21)(13,22)
\psline{-}(13,22)(14,21)

\psline{-}(0,25)(1,25)
\psline{-}(1,25)(2,28)
\psline{-}(2,28)(3,25)
\psline{-}(3,25)(4,28)
\psline{-}(4,28)(5,25)
\psline{-}(5,25)(6,28)
\psline{-}(6,28)(7,25)
\psline{-}(7,25)(8,28)
\psline{-}(8,28)(9,25)
\psline{-}(9,25)(10,28)
\psline{-}(10,28)(11,25)
\psline{-}(11,25)(12,28)
\psline{-}(12,28)(13,25)
\psline{-}(13,25)(14,28)

\psline{-}(0,30)(1,33)
\psline{-}(1,33)(2,30)
\psline{-}(2,30)(3,33)
\psline{-}(3,33)(4,30)
\psline{-}(4,30)(5,33)
\psline{-}(5,33)(6,30)
\psline{-}(6,30)(7,33)
\psline{-}(7,33)(8,30)
\psline{-}(8,30)(9,33)
\psline{-}(9,33)(10,30)
\psline{-}(10,30)(11,33)
\psline{-}(11,33)(12,30)
\psline{-}(12,30)(13,33)
\psline{-}(13,33)(14,30)

}
\end{pspicture}
\end{center}
\end{figure}
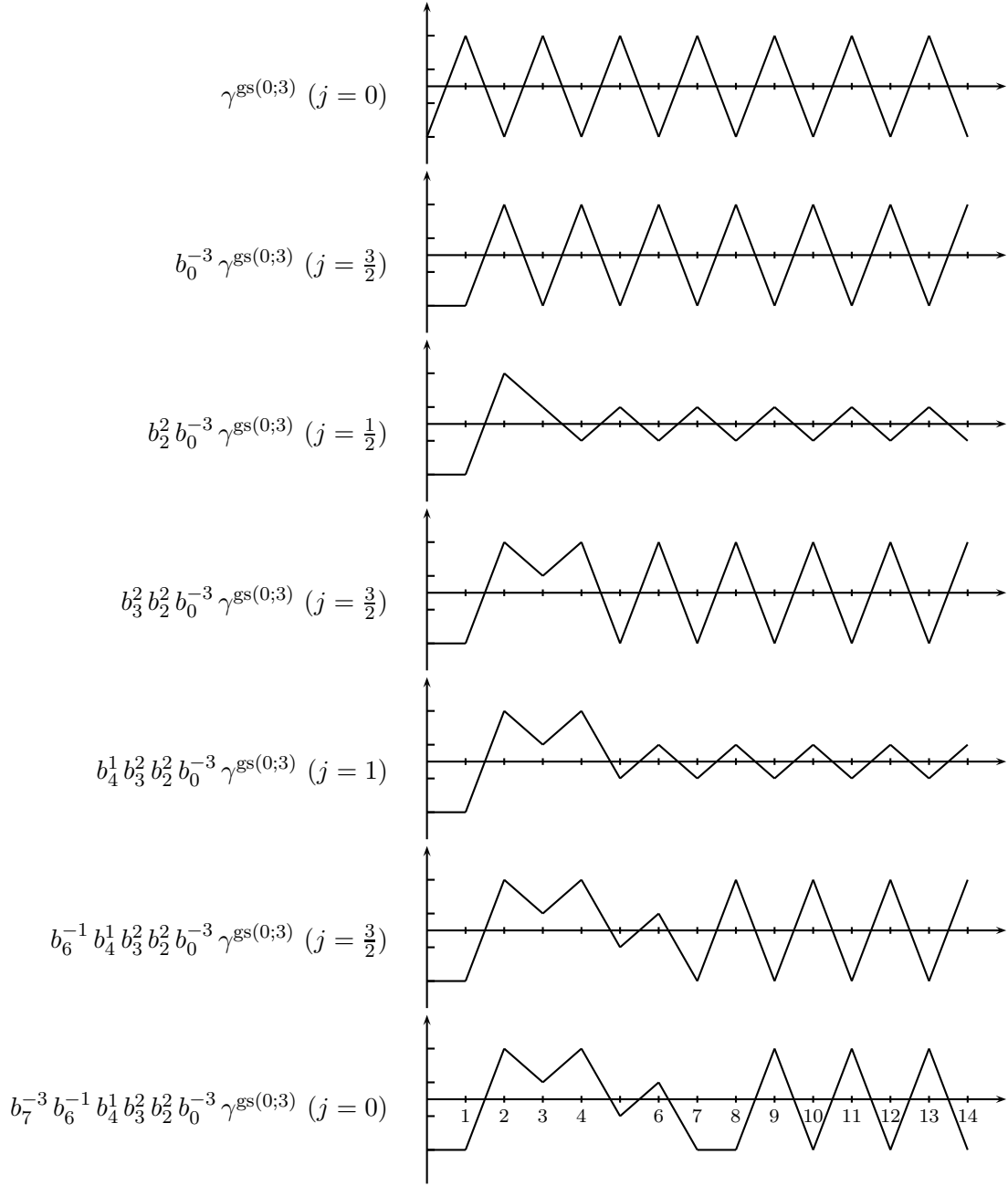

We illustrate this operator construction by re-expressing the $\mkp{3}$ path  of Fig. \ref{exemplemk} as a sequence of operators acting on $\ym^{\text{\gs}(0;3)}$. The first step is to construct the vector $\vec a=(a_0,a_1,\cdots )$ using (\ref{gavsa}). From (\ref{exemplemky}),
 we get
\begin{align}
	\vec a=(-3,0,2,2,1,0,-1,-3,0,0,...),
\label{exm3aa}
\end{align} 
which yields (cf. (\ref{pirev})):
\begin{align}
	\ym=\abp{7}{-3}~\abp{6}{-1}~\abp{4}{1}~\abp{3}{2}~\abp{2}{2}~\abp{0}{-3}~\ym^{\text{\gs}(0;3)}.\label{exm3}
\end{align}
The reconstruction of the path is worked out step by step in Fig. \ref{fig3}.

It is convenient to introduce a notation for a sequence of operators and its reversed version:
\begin{equation}
	\pi_b\equiv \abp{0}{a_0} \abp{1}{a_1}\abp{2}{a_2}\cdots 
\qquad\text{and}\qquad
	\pi_b^{\text{rev}}=\cdots \abp{2}{a_2} \abp{1}{a_1}\abp{0}{a_0},
	\end{equation}
so that $
	\ga=\pi_b^{\text{rev}}\ga^{\gs(0;k)}$. A path is completely defined by its associated string $\pi_b$. 
For the example (\ref{exm3}), we have
\begin{align}
	\pi_b^{\text{rev}}=\abp{7}{-3}~\abp{6}{-1}~\abp{4}{1}~\abp{3}{2}~\abp{2}{2}~\abp{0}{-3}\qquad\implies\qquad
	\pi_b= \abp{0}{-3} ~\abp{2}{2}~\abp{3}{2}~\abp{4}{1} ~\abp{6}{-1}~\abp{7}{-3}.
	\label{exm3a}
\end{align}
Of course the data $\pi_b$ and $\vec a$ are equivalent.
Fig. \ref{exS4} presents another example of the construction of the vector $\vec a$ from a path.

We denote the set of sequences $\pi_b$ generating $\mkp{k}$ paths by $\Pi^{(k)}$.  
The subset of sequences $\pi_b$ containing $n_+$ and $n_-$ operators $\ab$ and $\ab^{-1}$ respectively will be denoted by  $\Pi^{(k)}(n_+,n_-)$ and the set of corresponding paths by $\mkp{k}(n_+,n_-)$.

\subsection{Characteristics of an operator content}\label{carop}

Let us first express $j$ in terms of the $a_l$. This follows from the limiting form of $j_l$, which is itself readily obtained by  comparing the expression of $\ga_l$ in (\ref{gal=}) with the first equality in (\ref{defdejl}):
\begin{align}
j_{l}=\frac{1}{2}\sum_{n=0}^{l-1}(-1)^{n+1} a_n\qquad\implies \qquad j=\frac{1}{2}\sum_{n\geq0}(-1)^{n+1} a_n.\label{jb}
\end{align}
For instance, with $\ym^{\text{\gs}(j;k)}=\abp{0}{-2j} \ym^{\text{\gs}(0;k)}$, we see that $j_l(\ym^{\text{\gs}(j;k)})=j$ for all $l\geq 1$. 
For the example of Fig. \ref{fig3}, the corresponding sequence of values $j_l$ is  easily checked to be
\begin{align}\label{jlex}
	\jv=\left(\tfrac32,\tfrac32,\tfrac12,\tfrac32,1,1,\tfrac32,0,0,\cdots\right),
\end{align}
i.e, $j_{l\geq 8} =0$, while for all $n<8$, $0\leq j_n\leq \tfrac32$ as it should since $k=3$.

\begin{figure}[ht]\
\caption{Example of a $\mk^{(4)}_1$ path, with the vectors $\ga$ and $\vec a$ (whose entries are calculated from the half-sum of the one just above plus its right neighbor) written explicitly below. From the latter, we see that $n_+=18$ and $n_-=10$, so that $s=4$. (Ignore for the moment the vector $\vec h$ whose significance  will be explained in App. \ref{Nag}.) Here we have an example of a path for which $j=k/4$: for all those cases, the tail is a straight line on the axis. (Another example at level 2 is provided by Fig. 21 of \cite{LPM}.)}
\label{exS4}
\begin{center}
\begin{pspicture}(3,-1)(8,2.7)
{\psset{yunit=18pt,xunit=18pt,linewidth=.8pt}
\psline{->}(0,-0.5)(0,4.5) \psline{->}(0,2)(18,2)
\rput(2,1.4){{\scriptsize $2$}}
\rput(1,1.4){{\scriptsize $1$}}
\rput(3,1.4){{\scriptsize $3$}}
\rput(5,1.4){{\scriptsize $5$}}
\rput(7,1.4){{\scriptsize $7$}}
\rput(9,1.4){{\scriptsize $9$}}
\rput(11,1.4){{\scriptsize $11$}}
\rput(13,1.4){{\scriptsize $13$}}
\rput(4,1.4){{\scriptsize $4$}}
\rput(6,1.4){{\scriptsize $6$}}
\rput(8,1.4){{\scriptsize $8$}}
\rput(10,1.4){{\scriptsize $10$}}
\rput(12,1.4){{\scriptsize $12$}}
\rput(14,1.4){{\scriptsize $14$}}
\rput(15,1.4){{\scriptsize $15$}}
\rput(16,1.4){{\scriptsize $16$}}
\rput(17,1.4){{\scriptsize $17$}}

\rput[r](-0.2,2){{\scriptsize $0$}}
\rput[r](-0.2,0){{\scriptsize $-4$}}
\rput[r](-0.2,1){{\scriptsize $-2$}}
\rput[r](-0.2,3){{\scriptsize $2$}}
\rput[r](-0.2,4){{\scriptsize $4$}}

\psset{linestyle=solid}
\psline{-}(1,1.9)(1,2.1)\psline{-}(2,1.9)(2,2.1)
\psline{-}(3,1.9)(3,2.1)\psline{-}(4,1.9)(4,2.1)
\psline{-}(5,1.9)(5,2.1)\psline{-}(6,1.9)(6,2.1)
\psline{-}(7,1.9)(7,2.1)\psline{-}(8,1.9)(8,2.1)
\psline{-}(9,1.9)(9,2.1)\psline{-}(10,1.9)(10,2.1)
\psline{-}(11,1.9)(11,2.1)\psline{-}(12,1.9)(12,2.1)
\psline{-}(13,1.9)(13,2.1)\psline{-}(14,1.9)(14,2.1)
\psline{-}(15,1.9)(15,2.1)\psline{-}(16,1.9)(16,2.1)
\psline{-}(17,1.9)(17,2.1)

\psline{-}(0,1)(.1,1)
\psline{-}(0,2)(.1,2)
\psline{-}(0,3)(.1,3)
\psline{-}(0,0)(.1,0)
\psline{-}(0,4)(.1,4)

\psline{-}(0,0)(1,3)
\psline{-}(1,3)(2,3)
\psline{-}(2,3)(3,2)
\psline{-}(3,2)(4,0)
\psline{-}(4,0)(5,4)
\psline{-}(5,4)(6,3)
\psline{-}(6,3)(7,1)
\psline{-}(7,1)(8,3)
\psline{-}(8,3)(9,3)\psline{-}(9,3)(10,4)
\psline{-}(10,4)(11,0)\psline{-}(11,0)(12,0)
\psline{-}(12,0)(13,1)\psline{-}(13,1)(14,4)
\psline{-}(14,4)(15,3)\psline{-}(15,3)(16,3)
\psline{-}(16,3)(17,2)

\rput(-1,-1){{\scriptsize $\ga=$}}
\rput(0,-1){{\scriptsize $(-4$}}
\rput(1,-1){{\scriptsize $2$}}
\rput(2,-1){{\scriptsize $2$}}
\rput(3,-1){{\scriptsize $0$}}
\rput(4,-1){{\scriptsize $-4$}}
\rput(5,-1){{\scriptsize $4$}}
\rput(6,-1){{\scriptsize $2$}}
\rput(7,-1){{\scriptsize $-2$}}
\rput(8,-1){{\scriptsize $2$}}
\rput(9,-1){{\scriptsize $2$}}
\rput(10,-1){{\scriptsize $4$}}
\rput(11,-1){{\scriptsize $-4$}}
\rput(12,-1){{\scriptsize $-4$}}
\rput(13,-1){{\scriptsize $-2$}}
\rput(14,-1){{\scriptsize $4$}}
\rput(15,-1){{\scriptsize $2$}}
\rput(16,-1){{\scriptsize $2$}}
\rput(17,-1){{\scriptsize $0$}}
\rput(18,-1){{\scriptsize $\cdots)$}}

\rput(-1,-1.6){{\scriptsize $\vec a=$}}
\rput(0,-1.6){{\scriptsize $(-1$}}
\rput(1,-1.6){{\scriptsize $2$}}
\rput(2,-1.6){{\scriptsize $1$}}
\rput(3,-1.6){{\scriptsize $-2$}}
\rput(4,-1.6){{\scriptsize $0$}}
\rput(5,-1.6){{\scriptsize $3$}}
\rput(6,-1.6){{\scriptsize $0$}}
\rput(7,-1.6){{\scriptsize $0$}}
\rput(8,-1.6){{\scriptsize $2$}}
\rput(9,-1.6){{\scriptsize $3$}}
\rput(10,-1.6){{\scriptsize $0$}}
\rput(11,-1.6){{\scriptsize $-4$}}
\rput(12,-1.6){{\scriptsize $-3$}}
\rput(13,-1.6){{\scriptsize $1$}}
\rput(14,-1.6){{\scriptsize $3$}}
\rput(15,-1.6){{\scriptsize $2$}}
\rput(16,-1.6){{\scriptsize $1$}}
\rput(17,-1.6){{\scriptsize $0$}}
\rput(18,-1.6){{\scriptsize $\cdots)$}}

\rput(-1,-2.2){{\scriptsize $\vec h=$}}
\rput(0,-2.2){{\scriptsize $(\phantom{-}\underline4$}}
\rput(0.6,-2.2){{\scriptsize $|$}}
\rput(1,-2.2){{\scriptsize $\underline3$}}
\rput(1.6,-2.2){{\scriptsize $|$}}
\rput(2,-2.2){{\scriptsize $2$}}
\rput(3,-2.2){{\scriptsize $2$}}
\rput(3.6,-2.2){{\scriptsize $|$}}
\rput(4,-2.2){{\scriptsize $\underline4$}}
\rput(4.6,-2.2){{\scriptsize $|$}}
\rput(5,-2.2){{\scriptsize $\underline3$}}
\rput(6,-2.2){{\scriptsize $1$}}
\rput(7,-2.2){{\scriptsize $3$}}
\rput(7.6,-2.2){{\scriptsize $|$}}
\rput(8,-2.2){{\scriptsize $\underline3$}}
\rput(8.6,-2.2){{\scriptsize $|$}}
\rput(9,-2.2){{\scriptsize $\underline4$}}
\rput(10,-2.2){{\scriptsize $0$}}
\rput(11,-2.2){{\scriptsize $4$}}
\rput(11.6,-2.2){{\scriptsize $|$}}
\rput(12,-2.2){{\scriptsize $\underline4$}}
\rput(12.6,-2.2){{\scriptsize $|$}}
\rput(13,-2.2){{\scriptsize $\underline4$}}
\rput(13.6,-2.2){{\scriptsize $|$}}
\rput(14,-2.2){{\scriptsize $\underline3$}}
\rput(14.6,-2.2){{\scriptsize $|$}}
\rput(15,-2.2){{\scriptsize $\underline3$}}
\rput(15.6,-2.2){{\scriptsize $|$}}
\rput(16,-2.2){{\scriptsize $2$}}
\rput(17,-2.2){{\scriptsize $2$}}
\rput(18,-2.2){{\scriptsize $\cdots)$}}

}
\end{pspicture}
\end{center}
\end{figure}
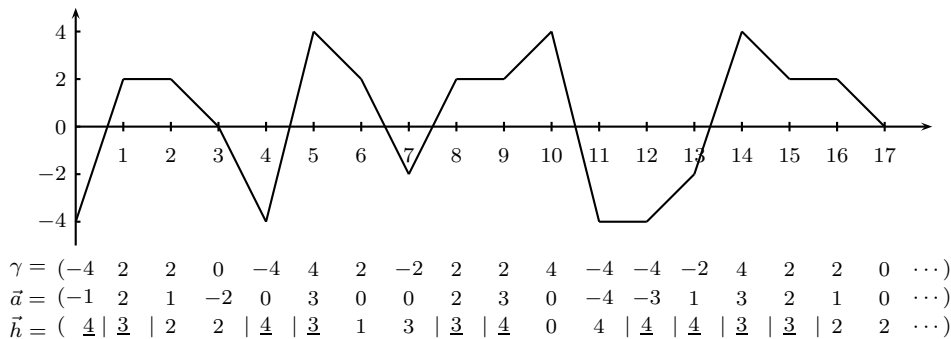

Let us next  express $s$  in terms of the $a_l$. Let $L$ be such that $a_l=0$ for $l\geq L$. Then
\begin{align}
	s&=-j+\frac12\sum_{l=1}^{L}(\ga_l-\ga_l^{\gs(j;k)})\nonumber\\
	&=-j+\frac14(\ga_1+\ga_L)+\frac14\sum_{l=1}^{L-1}(\ga_l+\ga_{l+1})\nonumber\\
	&\phantom{=-j}\;-\frac14(\ga_1^{\gs(j;k)}+\ga_L^{\gs(j;k)})-\frac14\sum_{l=1}^{L-1}(\ga_l^{\gs(j;k)}+\ga_{l+1}^{\gs(j;k)})\nonumber\\
&=-j+\frac14(k+2a_0)+\frac12\sum_{l\geq1}a_l-\frac14(k-4j)\nonumber\\	
&=	\frac12\sum_{l\geq 0}a_l\label{cb}.
\end{align}
In the third step, we used $\ga_L=\ga_L^{\gs(j;k)}$ and the Eqs. (\ref{gal=}), (\ref{gavsa}) and (\ref{groundk}).
With $n_+$ and $n_-$ defined respectively as the sum of positive and negative $a_l$:
\begin{align}\label{nn}
	n_+=\sum_{l\geq0,\,a_l>0}|a_l|,\qquad
	n_-=\sum_{l\geq0,\,a_l<0}|a_l|,
\end{align}
the sector can then be rewritten as
\begin{align}\label{secgk}
	s=\frac12(n_+-n_-).
\end{align}
This formula is illustrated in Fig. \ref{exS4}.

We next derive the expression of the weight $w(\ga)$ in terms of the $a_l$. Using (\ref{gavsa}), the partial weight $w_l$ defined in (\ref{weightak}) is:
\begin{align}
	\wap{l}&=\frac{l}4|\ga_l+\ga_{l+1}|=\frac{l}{2}|a_l|\qquad \implies \qquad\sum_{l\geq1} w_l=\frac12\sum_{l\geq1} l|a_l| \label{wb}.
\end{align}
For the example of Fig. \ref{exemplemk}, we find $n_+=5$ and $n_-=7$, so the sector is $-1$. From the tail, $j=0$, so that the weight computed from (\ref{wb}) is:
\begin{align}
	\wa=-\frac12(-1+0)+\frac12(4+6+4+6+21)=21.
\end{align}
Of course, the part $-(s+j)/2$ can also be expressed in terms of $a_l$, resulting into the compact expression
\begin{align}
2\wa(\ga)=\sum_{l\geq1} l|a_l|-\sum_{l\geq1, \,l \,\text{odd}}a_l. 
\label{wba}
\end{align}

Finally, we express the original weight $\wm(\ga)$ also in terms of the $a_l$, thereby ending up with a simple proof of 
 the equivalence between the weights (\ref{weightmk}) and (\ref{weightak}).
Using (\ref{gavsa}),
 we first observe that 
 \begin{align}
\text{min}(\gamma_l,-\gamma_{l+1})
       &=\text{min}(\gamma_l,\gamma_l-2a_l)=  \gamma_l-2a_l\, \theta(a_l)\label{tre1}
\end{align}
where $\theta(x)$ is the step function. On the other hand,
  $\text{min}(\gamma^{\gs}_l,-\gamma^{\gs}_{l+1})= \gamma^{\gs}_l $
    since the two terms are equal.
Therefore, from (\ref{weightmk}) and the simple identity 
\begin{align}
\sum_{n>0}ng_n=\sum_{n>0}\lfloor(n+1)/2\rfloor(g_n+g_{n+1}),  \end{align} 
 we can write
\begin{align}2 \hat w (\gamma)
    &= - \sum_{l>0}  l (\gamma_l - \gamma^{\gs}_l)
             + 2 \sum_{l>0,\,a_l>0} l a_l\nonumber\\ &
    = - \sum_{l>0}  \lfloor(l+1)/2\rfloor (\gamma_l + \gamma_{l+1}
                               - \gamma^{\gs}_l - \gamma^{\gs}_{l+1})
             + 2 \sum_{l>0,\, a_l>0} l a_l\nonumber\\
  &  = - \sum_{l>0}  \lfloor(l+1)/2\rfloor (\gamma_l + \gamma_{l+1})
             + 2 \sum_{l>0,\,a_l>0} l a_l,\label{tre2}
             \end{align}
since $\gamma^{\gs}_l=-\gamma^{\gs}_{l+1}$. Using (\ref{gavsa}) again, we obtain
\begin{align}2 \hat w (\gamma)
&  = - 2 \sum_{l>0}  \lfloor(l+1)/2\rfloor a_l
             + 2 \sum_{l>0,\,a_l>0} l a_l
    \nonumber\\& = - \sum_{l>0, \,l \,\text{even}}  l a_l - \sum_{l>0,\,l\,\text{odd}} (l+1) a_l
             + 2 \sum_{l>0,\,a_l>0} l a_l
    \nonumber\\& = - \sum_{l>0}  l a_l - \sum_{l>0, \,l \,\text{odd}} a_l
             + 2 \sum_{l>0,\, a_l>0} l a_l
    \nonumber\\&= \sum_{l>0}  l |a_l| - \sum_{l>0,\, l \,\text{odd}} a_l \nonumber\\&=2w(\ga),\label{tre3}
\end{align}
as announced.

\section{Strategy for the construction of the $\su{k}$ paths generating functions: restricted sequences of operators and operators acting on them}\label{Strategy}

The aim of introducing the operator representation of paths is to set up a constructive method for obtaining their generating function (see e.g., \cite{JM,JMsusy,LPM}).\footnote{This has also been used for constructing a bijective proof between two different path formulations in \cite{BMW}.} 
The problem of summing over all paths $\ga\in\mkp{k}$ in (\ref{cagf}) is transformed into that of summing over all $\pi_b\in\Pi^{(k)}$, which is more manageable.

 In principle, given this operator construction, we should be in position to work out the generating function by following the standard strategy \cite{OleJS,PMjmp} used in the level-1 case \cite{LPM}, namely, $q$-enumerate sequences of operators 
 from a minimal-weight configuration for a fixed operator content, out of which all other configurations are generated by operator displacements, followed by a summation over the number of operators of each type. However, a frontal attack along this line  proves to be difficult. We thus proceed along a different route and  break the problem in two parts.

\subsection{Restricted sequences}\label{opb}

 In the first step, we consider only the subclass of paths $\mkp{k}_{\text{res}}\subset\mkp{k}$ where $\ga^0\in\mkp{k}_{\text{res}}$ is of the form 
\begin{align}\label{ga0}
\ga^0=(\pi_b^0)^{\text{rev}}\ym^{\gs(0;k)}\qquad\text{where}\qquad \pi_b^0\in\Pi^{(1)}.
\end{align}
Such $\pi_b^0$ will be called restricted sequences. Note in particular that a restricted sequence is associated to a vector $\vec a$ such that $|a_l|\leq1$ for every $l$. But actually it must  satisfy the stronger repulsion conditions (cf. \cite[Eq. (61)]{LPM})
\begin{align}
\label{succ}
	\abp{l'}{\pm1}\abp{l}{\mp1}\ne0\; \Leftrightarrow \;l'-l\,> 0\;  \text{and even} \qquad\text{and}\qquad
	\abp{l'}{\pm1}\abp{l}{\pm1}\ne0\;\Leftrightarrow \; l'-l\,>0\;  \text{and odd},
\end{align}
(which force $|a_l|\leq1$).
The requirement (\ref{succ}) is equivalent to the statement that for $\ga^0\in\mkp{k}_{\text{res}}$,  all $j_l$ are in the range $0\leq j_l\leq 1/2$. That $|a_l|\leq1$ is  not sufficient to ensure this can be seen from the example
\begin{align}\pi'_b=b_0^{-1}b_2^{-1}b_3^{1}\qquad\implies \qquad \vec j =(\tfrac12,\tfrac12,1,\tfrac32,\tfrac32,\cdots),
\end{align}
a sequence for which the two conditions in (\ref{succ}) are violated (so that $\pi'_b\not\in\Pi^{(1)}$).
(For other examples, see Table \ref{tab1} in App. \ref{Nag}.)

 We stress that while these paths (\ref{ga0}) obtained from a restricted sequence at level $k$ have the same operator content -- and therefore the same values of  $\wa$, $\cm$ and $j$ -- as their $k=1$ image, they are not identical paths: indeed, the operator sequence act on  different ground states. For example, consider the sequence 
$\pi_b^0=b_0^{-1}~b_2^{1}~b_3^{1}~b_6^{1}$ and the associated path for $k=1$ 
\begin{align}
\ym^{\gs(0;1)}=(-1,1,-1,1,-1,\cdots)\quad \Rightarrow	\quad (\pi_b^0)^{\text{rev}}\ym^{\gs(0;1)}=
(-1,-1,1,1,1,-1,1,1,-1,\cdots),
\end{align}
and for $k=3$
\begin{align}
\ym^{\gs(0;3)}=(-2,3,-3,3,-3,\cdots)\quad \Rightarrow	\quad (\pi_b^0)^{\text{rev}}\ym^{\gs(0;3)}=
(-3,1,-1,3,-1,1,-1,3,-3,\cdots).
\end{align}
For another example, see Fig. \ref{expi0} below. However, their relationship is clearly one-to-one: both sets $\mkp{1}$ and  $\mkp{k}_{\text{res}}$ are bijectively related to the set $\Pi^{(1)}$.

Let $G_0(n_+,n_-;z;q)$ be the generating function for paths $\ga^0\in\mkp{k}_{\text{res}}(n_+,n_-)$:
\begin{align}
	G_0(n_+,n_-;z;q)=\sum_{\ga^0\in\mkp{k}_{\text{res}}(n_+,n_-)}q^{\wt(\ga^0)}z^{s(\ga^0)},
		\label{chark1d}
\end{align}
where we define
\begin{equation}\label{defwt}
\wt(\ga) =w(\ga)+\frac12j.
\end{equation}
This generating function is that for level-1 paths without the $-j/2$ factor \cite[Eq. (75)]{LPM}:
\begin{align}
	G_0(n_+,n_-;z;q)=\frac{q^{\frac{1}{4}(n_++n_-)^2}z^{{\frac{1}{2}(n_+-n_-)}}}{(q)_{n_+}(q)_{n_-}}.
		\label{chark1}
\end{align}
At this point we do not include the correction $j/2$ to the weight as the value of $j$ will vary in the subsequent stage of the construction, to which we now turn.
 

\subsection{The action of $c_i$  on $\pi_b$}\label{operateursci}

In the second step, we generate generic sequences out of restricted ones.
For this construction, we introduce a new type of operator, $c_i$, whose action is defined directly on  
elements of $\Pi^{(k)}(n_+,n_-)$  (and note that $\Pi^{(1)}\subset\Pi^{(k)}$).
This action keeps the number of operators $n_\pm$ constant. As a result, it does not modify the sector, which is thus fixed by that of $\pi_b^0$, as $(n_+-n_-)/2$.

The acting position $i$  of the operator $c_i$  will refer to the position of the first $b^{\pm1}$ operator acted  on (this action affects more than one operator).  
This position, read from left to  right in $\pi_b$, is called the rank. For instance, the operator $b_6^{-1}$ in the sequence $\pi_b$ of Eq. (\ref{exm3a}) has rank 9. 
The position $l_i$ of the operator of rank $i$ is defined as follows. 
Transform the vector $\vec a$ by replacing each entry $a_n$ by the $|a_n|$-tuple $n,\cdots,n$, that is,
\begin{equation}
(a_0,a_1,\cdots, a_n,\cdots)\to (\underbrace{0,\cdots,0}_{|a_0|},\underbrace{1,\cdots,1}_{|a_1|},\cdots,\underbrace{n,\cdots,n}_{|a_n|},\cdots)
\end{equation}
(vanishing $a_i$ do not contribute) and underline each entry $n$ if $a_n<0$. Call the resulting 
 vector $\vec l$. For $\pi_b\in\Pi^{(k)}(n_+,n_-)$, this vector as $n_++n_-$ entries:\footnote{A more compact but less transparent definition is:
$	l_i=\text{min}\,\left\{l\,\big|	\, \sum_{l'=0}^{l}|a_{l'}|\geq i\right\}.$}
\begin{equation}
\vec l=(l_1,l_2,...,l_{n_++n_-}),
\end{equation}
$n_-$ of which are underlined. 
For instance, from the vector $\vec a$ of the example (\ref{exm3aa}), we find:
\begin{align}\label{exl}
	\vec l=(\underline{0},\underline{0},\underline{0},2,2,3,3,4,\underline{6},\underline{7},\underline{7 },\underline{7}).
\end{align}
(We stress that $\pi_b, \vec a$ and $\vec l$ are three different but equivalent  ways of coding the path.)


With $l_i$  defined, we are now in position to describe the action of $c_i$. We first impose an ordering: the operators $c_i$ must act successively in strictly increasing values of $i$ on some element $\pi_b^0\in\Pi^{(1)}(n_+,n_-)$. Let $\pi_b$ be a sequence obtained from the action of a number of $c_j$ with $j<i$ and set $c_i\pi_b=\pi_b'$. Then $\pi_b'$ is obtained by decreasing by one the position at which all operators of $\pi_b$ of rank $>i$ act, that is, to replace all $b_l^{a_l}$ with $l> l_i$  by $b_{l-1}^{a_l}$. It is understood that this action is non-zero  provided the resulting sequence is allowed, meaning that the components of the new vector $\vec a'$ satisfy (\ref{cona}), and the operator content is unchanged. Phrased differently, 
if $\pi_b=\prod_{n}  b_n^{a_n} $, we have
\begin{align}\label{cisurpi}
	c_i\,\pi_b=\pi'_b=
		\begin{cases}\prod_{0\leq m\leq l_i} b_m^{a_m} \prod_{n> l_i} b_{n-1}^{a_n} 
		\qquad &\mbox{if} \;\pi_b'\in\Pi^{(k)}(n_+,n_-)\\		0\qquad &\mbox{if} \;\pi_b'\not \in\Pi^{(k)}(n_+,n_-).\end{cases}
\end{align}
We stress that the ordering condition on the action of the $c_i$ 
 ensures that
 the conditions (\ref{succ}) are satisfied for all operators $b^{\pm1}$ of rank $\geq i$ in $\pi_b$.

Alternatively, we can define the action of $c_i$ (when non-zero) on the $\vec l$ vector to be: 
\begin{equation}\label{cionl}
	c_i(l_1,...,l_i,l_{i+1},l_{i+2},...,l_{n_++n_-})=
				(l_1,...,l_i,l_{i+1}-1,l_{i+2}-1,...,l_{n_++n_-}-1).
\end{equation}
Notice that if $l_{i+1}=l_{i}+1$, since the sequence $\pi_b$  on which $c_i$ acts is such  that $a_{l_{i+1}}=\pm 1$, then
after the action of $c_i$, $a_{l_{i}} \to a_{l_{i}}\pm1$, that is
\begin{align}\label{increasing al}
	c_i:\,\cdots\abp{l_{i}}{a_{l_{i}}}\abp{l_{i}+1}{\pm1}\cdots\rw\cdots\abp{l_{i}}{a_{l_{i}}\pm1}\cdots
		\qquad(\text{when}\quad \text{sgn}(a_{l_i})=\pm1).
	\end{align}
This is thus precisely how the absolute values of the various $a_l$ can be increased successively. Note that the action of $c_i$ is clearly nonlocal.

 We denote an admissible sequence of $c_i$ by $\pi_c$, with 
\begin{equation}
\pi_b=\pi_c^{\text{rev}}~\pi_b^0.
\end{equation}
By admissible we mean that no operator within the sequence acts trivially (i.e., whose action yields 0).

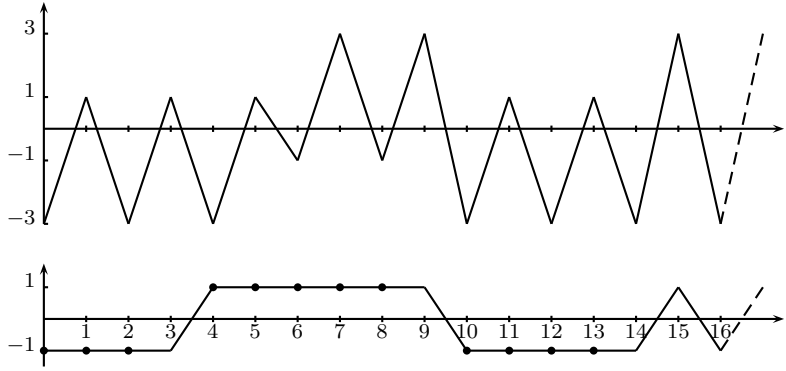
\begin{figure}[ht]\
\caption{The restricted  path  
$(\pi_b^0)^{\text{rev}}\ga^{\gs(0;3)}\in \mkp{3}_{\text{res}}$  
associated to the sequence $\pi^0_b$ given in (\ref{excsurba}). Below, we have displayed the corresponding level-1 path $(\pi_b^0)^{\text{rev}}\ga^{\gs(0;1)}$. In the latter case, we have indicated by
dots the application points of the operators $b^{\pm1}$. These correspond to the positions of the particles. (We recall that the particles are horizontal edges at $\pm1$ and their position is the edge initial point -- cf. the description of \cite[Section 2.4.3]{LPM}.) In contrast, the particle content  is not very transparent in the level-3 restricted path.}
\label{expi0}
\begin{center}
\begin{pspicture}(0,-0.5)(9,4.5)
{\psset{yunit=12pt,xunit=16pt,linewidth=.8pt}

\psline{->}(0,3)(0,10) \psline{->}(0,6)(17.5,6)

\rput[Br](-0.2,3){{\scriptsize $-3$}}
\rput[Br](-0.2,5){{\scriptsize $-1$}}
\rput[Br](-0.2,7){{\scriptsize $1$}}
\rput[Br](-0.2,9){{\scriptsize $3$}}

\psset{linestyle=solid}
\psline{-}(1,5.9)(1,6.1)\psline{-}(2,5.9)(2,6.1)
\psline{-}(3,5.9)(3,6.1)\psline{-}(4,5.9)(4,6.1)
\psline{-}(5,5.9)(5,6.1)\psline{-}(6,5.9)(6,6.1)
\psline{-}(7,5.9)(7,6.1)\psline{-}(8,5.9)(8,6.1)
\psline{-}(9,5.9)(9,6.1)\psline{-}(10,5.9)(10,6.1)
\psline{-}(11,5.9)(11,6.1)
\psline{-}(12,5.9)(12,6.1)
\psline{-}(13,5.9)(13,6.1)
\psline{-}(14,5.9)(14,6.1)
\psline{-}(15,5.9)(15,6.1)\psline{-}(16,5.9)(16,6.1)

\psline{-}(0,3)(1,7)
\psline{-}(1,7)(2,3)
\psline{-}(2,3)(3,7)
\psline{-}(3,7)(4,3)
\psline{-}(4,3)(5,7)
\psline{-}(5,7)(6,5)
\psline{-}(6,5)(7,9)
\psline{-}(7,9)(8,5)
\psline{-}(8,5)(9,9)
\psline{-}(9,9)(10,3)
\psline{-}(10,3)(11,7)
\psline{-}(11,7)(12,3)
\psline{-}(12,3)(13,7)
\psline{-}(13,7)(14,3)
\psline{-}(14,3)(15,9)
\psline{-}(15,9)(16,3)

\psline{-}(0,3)(0.1,3)
\psline{-}(0,5)(0.1,5)
\psline{-}(0,7)(0.1,7)
\psline{-}(0,9)(0.1,9)

\psline{->}(0,-1.5)(0,1.75) \psline{->}(0,0)(17.5,0)
\rput(2,-0.4){{\scriptsize $2$}}
\rput(3,-0.4){{\scriptsize $3$}}
\rput(1,-0.4){{\scriptsize $1$}}
\rput(5,-0.4){{\scriptsize $5$}}
\rput(7,-0.4){{\scriptsize $7$}}
\rput(9,-0.4){{\scriptsize $9$}}
\rput(4,-0.4){{\scriptsize $4$}}
\rput(6,-0.4){{\scriptsize $6$}}
\rput(8,-0.4){{\scriptsize $8$}}
\rput(10,-0.4){{\scriptsize $10$}}
\rput(11,-0.4){{\scriptsize $11$}}
\rput(12,-0.4){{\scriptsize $12$}}
\rput(13,-0.4){{\scriptsize $13$}}
\rput(14,-0.4){{\scriptsize $14$}}
\rput(15,-0.4){{\scriptsize $15$}}
\rput(16,-0.4){{\scriptsize $16$}}

\psset{dotsize=3pt}\psset{dotstyle=*}
\psdots(0,-1)(1,-1)(2,-1)(4,1)(5,1)(6,1)(7,1)(8,1)(10,-1)(11,-1)(12,-1)(13,-1)

\rput[Br](-0.2,-1){{\scriptsize $-1$}}
\rput[Br](-0.2,1){{\scriptsize $1$}}

\psset{linestyle=solid}
\psline{-}(1,-0.1)(1,0.1)\psline{-}(2,-0.1)(2,0.1)
\psline{-}(3,-0.1)(3,0.1)\psline{-}(4,-0.1)(4,0.1)
\psline{-}(5,-0.1)(5,0.1)\psline{-}(6,-0.1)(6,0.1)
\psline{-}(7,-0.1)(7,0.1)\psline{-}(8,-0.1)(8,0.1)
\psline{-}(9,-0.1)(9,0.1)\psline{-}(10,-0.1)(10,0.1)
\psline{-}(11,-0.1)(11,0.1)
\psline{-}(12,-0.1)(12,0.1)
\psline{-}(13,-0.1)(13,0.1)
\psline{-}(14,-0.1)(14,0.1)
\psline{-}(15,-0.1)(15,0.1)
\psline{-}(16,-0.1)(16,0.1)

\psline{-}(0,-1)(0.1,-1)
\psline{-}(0,1)(0.1,1)

\psline{-}(0,-1)(3,-1)
\psline{-}(3,-1)(4,1)
\psline{-}(4,1)(9,1)
\psline{-}(9,1)(10,-1)
\psline{-}(10,-1)(14,-1)
\psline{-}(14,-1)(15,1)
\psline{-}(15,1)(16,-1)

\psset{linestyle=dashed}
\psline{-}(16,-1)(17,1)
\psline{-}(16,3)(17,9)
}

\end{pspicture}
\end{center}
\end{figure}

Let us illustrate the building up of the path $\ga$ given in (\ref{exm3}) -- or the corresponding sequence $\pi_b$ in (\ref{exm3a}) -- in terms of a string of  $\cip{i}$ acting on some $\pi_b^0$.
The required sequence $\pi_b^0$ is 
 (see Section \ref{generchemins} for the precise procedure for obtaining $\pi_b^0$ from $\pi_b$) 
\begin{align}\label{excsurba}
	\pi_b^0=\abp{0}{-1}\,\abp{1}{-1}\,\abp{2}{-1}\,\abp{4}{1}\,\abp{5}{1}
		\,\abp{6}{1}\,\abp{7}{1}\,\abp{8}{1}\,\abp{10}{-1}\,\abp{11}{-1}\,\abp{12}{-1}\,\abp{13}{-1} ,
\end{align}
whose weight is 40 ($s=-1$ and $j=0$). The corresponding restricted path $\in\mkp{3}_{\text{res}}$ as well as its level-1 version are pictured in Fig. \ref{expi0}. We then apply on this $\pi^0_b$ the suitable sequence of operators $c$ that modifies the powers of $b$ successively by steps of $\pm1$ each time:
\begin{align}\label{excsurb}
	\cip{1}\,\pi_b^0&=\abp{0}{-2}\,\abp{1}{-1}\,\abp{3}{1}\,\abp{4}{1}\,\abp{5}{1}\,\abp{6}{1}\,
		\abp{7}{1}\,\abp{9}{-1}\,\abp{10}{-1}\,\abp{11}{-1}\,\abp{12}{-1},&(\tfrac{68}2)\nonumber\\
	\cip{2}\,\cip{1}\,\pi_b^0&=\abp{0}{-3}\,\abp{2}{1}\,\abp{3}{1}\,\abp{4}{1}\,\abp{5}{1}\,\abp{6}{1}\,
		\abp{8}{-1}\,\abp{9}{-1}\,\abp{10}{-1}\,\abp{11}{-1},&(\tfrac{58}2)\nonumber\\
	\cip{4}\,\cip{2}\,\cip{1}\,\pi_b^0&=\abp{0}{-3}\,\abp{2}{2}\,\abp{3}{1}\,\abp{4}{1}\,\abp{5}{1}\,
		\abp{7}{-1}\,\abp{8}{-1}\,\abp{9}{-1}\,\abp{10}{-1},&(\tfrac{50}2)\nonumber\\
	\cip{6}\,\cip{4}\,\cip{2}\,\cip{1}\,\pi_b^0&=\abp{0}{-3}\,\abp{2}{2}\,\abp{3}{2}\,\abp{4}{1}\,\abp{6}{-1}\,
		\abp{7}{-1}\,\abp{8}{-1}\,\abp{9}{-1},&(\tfrac{44}2)\nonumber\\
	\cip{10}\,\cip{6}\,\cip{4}\,\cip{2}\,\cip{1}\,\pi_b^0&=\abp{0}{-3}\,\abp{2}{2}\,\abp{3}{2}\,
		\abp{4}{1}\,\abp{6}{-1}\,\abp{7}{-2}\,\abp{8}{-1},&(\tfrac{42}2)\nonumber\\
	\cip{11}\,\cip{10}\,\cip{6}\,\cip{4}\,\cip{2}\,\cip{1}\,\pi_b^0&=\abp{0}{-3}\,\abp{2}{2}\,\abp{3}{2}\,
		\abp{4}{1}\,\abp{6}{-1}\,\abp{7}{-3}&(\tfrac12+\tfrac{41}2).
\end{align}
At the right of each sequence, we have indicated the weight in the form (\ref{wb}). The paths corresponding to all these sequences all have $s=-1$ (according to (\ref{secgk}) and because $c_i$  does not modify the number of operators of each type) 
and $j=1$ except for the last one for which $j=0$. For this example, the sequence $\pi_c$ is thus $\pi_c=c_1c_2c_4c_6c_{10}c_{11}$.

Given that $c_i$ shifts by $-1$ the action position of all the operators $b$ of rank $> i$, it follows from the expression (\ref{wb}) that the effect of $c_i$ on the $a_l$ sum-part of the weight is
\begin{align}
	c_i:\frac12\sum_l l|a_l]&\to \frac12 \sum_{l\leq l_i} l|a_l| +\frac12 \sum_{l>l_i} (l-1)|a_l|\nonumber\\
	&=  \frac12 \sum_l l|a_l| -\tfrac12(n_++n_- -i)\label{poidsc}
\end{align}
using the constraints that $|a_l|=1$ for $l>l_i$.:
Therefore, with $\ga=\pi_b^{\text{rev}}\ga^{\gs(0;k)}$ and since $s$ is not changed by $c_i$, we have
\begin{equation}\label{wec}
\wt(c_i\ga)-\wt(\ga)=  -\tfrac12(n_++n_- -i),
\end{equation}
recalling the definition (\ref{defwt}) for $\wt$.
Note that $c_i$ also modifies $j$, hence the term $-j/2$ in the full weight expression (\ref{wb}), but this will be treated separately.

\subsection{Reduced vectors}\label{Sjh}

A key tool in the combinatorial analysis of the sequences generated by the $c_i$ on $\pi_b^0$ is the reformulation of the action of $c_i$ on a sort of deformed and reduced version of the vector $\jv$, which we call $\jh$. This reduced vector $\jh$ has $n_++n_-+1$ entries, the first of which is trivial, being set equal to 0. It is defined as: 
\begin{equation}\label{jhdef}
	\jh=(\jh_0,\jh_1,\cdots, \jh_i,\cdots,\jh_{n_++n_-}),
\end{equation}
where $\jh_i$ corresponds to the value of $j$ 
after the action of the $b^{\pm1}$ operator of rank $i$ at position $l_i$. By this definition, $\jh_0$ is the value before the application of any operator and is thus always 0. The other values $\jh_i$ are fixed recursively form the entries of the vector $\vec l$ as:
\begin{equation}\label{jil}
\jh_i-\jh_{i-1}=\frac12{\e_i}(-1)^{l_i-1},
\end{equation}
where $\e_i=-1$ if $l_i$ is underlined and 1 otherwise.
 For instance, to the sequence $\pi_b$ in (\ref{exm3a})
corresponds the  vector $\vec l$ of (\ref{exl}) and the reduced vector
\begin{equation}\label{jhex}
	\jh(\pi_b)= \l(0,\tfrac12,1,\tfrac32,1,\tfrac12,1,\tfrac32,1,\tfrac32,1,\tfrac12,0\r).
\end{equation}

It is easy to see that irrespective of the positions and the relative order of the operators $b^{\pm1}$ (in respective number $n_\pm$) in $\pi^0_b$, this vector $\jh$ is always of the form
\begin{equation}\label{jbase}
	\jh(\pi_b^0) \equiv\jh^{(0)}=\l(0,\tfrac12,0,\tfrac12,0,\tfrac12,0,\cdots,
		\tfrac12[n_++n_-]_2\r),
\end{equation}
using the notation  $[n]_2=\frac12(1-(-1)^n)$.
This structure is a direct consequence of (\ref{jil}), which implies $|\jh_i-\jh_{i-1}|=\frac12$, and the bound $0\leq\jh_i\leq\frac12$ that reflects the condition $\pi_b^0\in\Pi^{(1)}$.

 The relation (\ref{jbase}) demonstrates the advantage of reformulating the $q$-enumeration of sequences of $c_i$ on $\pi_b^0$ into that of $q$-enumerating sequences of $c_i$ on $\jh^{(0)}$: $\jh^{(0)}$ is unique. This reformulation isolates perfectly the objects we want to enumerate.

\subsection{The action  of $c_i$ on reduced vectors}\label{cisurjh}

To find the action of $c_i$ on a generic $\jh$, we recall from Eq. (\ref{jb}) that every operator $b_{l_i+1}^{\pm1}$ in a string $\pi_b$ acting on $\ga^{\gs(0;k)}$ modifies the value of $j_{l}$ for $l>l_i+1$ by $\frac12(-1)^{l_i}$. Hence, if $l_i+1$ is reduced by one, which is the effect of applying $c_i$, the sign of this modification term  of $j_l$ is inverted: $\frac12(-1)^{l_i}\to\frac12(-1)^{l_i-1}$. Thus, if the entries $\jh_t$ for $t>i-1$  oscillate between $\jh_{i}$ and $\jh_{i}\pm\frac12$   $(\equiv \jh_{i+1})$, after the action of $c_i$  they  oscillate between $\jh_{i}$ and $\jh_{i}\mp1/2$. 
The action of $c_i$ on $\jh$ is thus defined as follow:
\begin{align}\label{conj2}
c_i(\jh_0,\jh_1,\cdots,\jh_i,\jh_i\pm\tfrac12,\jh_i,\jh_i\pm\tfrac12,\cdots)=
(\jh_0,\jh_1,\cdots,\jh_i,\jh_i\mp\tfrac12,\jh_i,\jh_i\mp\tfrac12,\cdots),
\end{align}
and it vanishes if $\jh_i\mp\tfrac12$ is either negative or $>k/2.$
Equivalently, when non-zero, this action is
\begin{align}\label{conj}
	c_i\jh=\jh'
\quad\text{
where}\quad
	\jh'_t=
\begin{cases}		\jh_t&\text{for }t\leq i\text{ and }t=i+2n\\
		\jh_t \mp1&\text{for }t=i+2n+1\text{ if }\jh_n=\jh_i\mp\frac12[n-i]_2\text{ for }n\geq i-1\end{cases}
\end{align}
Since $j=\jh_{n_++n_-}$, it readily follows that under the action of $c_i$, $j$ changes by $\pm1$ or $0$.

To illustrate the formula (\ref{conj}), let us redo the example (\ref{excsurb}) in the $\jh$ language, thereby reconstructing the vector (\ref{jhex}) out of $\jh^{(0)}$: 
\begin{align}\label{construction jh}
	\jh^{(0)}&=
		\l(0,\tfrac12,0,\tfrac12,0,\tfrac12,0,\tfrac12,0,\tfrac12,0,\tfrac12,0\r),\nonumber\\
	\cip{1}\,\jh^{(0)}&=
		\l(0,\tfrac12,1,\tfrac12,1,\tfrac12,1,\tfrac12,1,\tfrac12,1,\tfrac12,1\r),\nonumber\\
	\cip{2}\,\cip{1}\,\jh^{(0)}&=
		\l(0,\tfrac12,1,\tfrac32,1,\tfrac32,1,\tfrac32,1,\tfrac32,1,\tfrac32,1\r),\nonumber\\
	\cip{4}\,\cip{2}\,\cip{1}\,\jh^{(0)}&=
		\l(0,\tfrac12,1,\tfrac32,1,\tfrac12,1,\tfrac12,1,\tfrac12,1,\tfrac12,1\r),\nonumber\\
	\cip{6}\,\cip{4}\,\cip{2}\,\cip{1}\,\jh^{(0)}&=
		\l(0,\tfrac12,1,\tfrac32,1,\tfrac12,1,\tfrac32,1,\tfrac32,1,\tfrac32,1\r),\nonumber\\
	\cip{10}\,\cip{6}\,\cip{4}\,\cip{2}\,\cip{1}\,\jh^{(0)}&=
		\l(0,\tfrac12,1,\tfrac32,1,\tfrac12,1,\tfrac32,1,\tfrac32,1,\tfrac12,1\r),\nonumber\\
	\cip{11}\,\cip{10}\,\cip{6}\,\cip{4}\,\cip{2}\,\cip{1}\,\jh^{(0)}&=
		\l(0,\tfrac12,1,\tfrac32,1,\tfrac12,1,\tfrac32,1,\tfrac32,1,\tfrac12,0\r).
\end{align}
\

\subsection{Reversing the construction}\label{generchemins}

Our construction consists thus in representing a path $\mkp{k}$ by a sequence $\pi_b\in\Pi^{(k)}$ and to generate the latter by a restricted sequence $\pi_b^0\in \Pi^{(1)} $ on which we act with a (reverted) sequence $\pi_c$ of operators $c_i$. 
That the pair $(\pi_c,\pi_b^{0})$ leads to a unique $\pi_b$, hence a unique path $\ga$, follows by construction. To prove that the relation $(\pi_c,\pi_b^{0})\lrw\pi_b$ is bijective, it suffices to show that the reverse   operation is
well defined.

Reversing the construction consists in deconstructing the sequence $\pi_b$, by undoing the $c_i$ actions on $\pi_b$ one by one, in decreasing values of the rank, until we reach the sequence $\pi_b^0$. 
To implement this deconstruction process, we need to determine, within $\pi_b$, where a $c_i$ has acted. Recall that to $\pi_b$ there corresponds a unique vector $\vec l$, hence a unique $\jh$. 
The $c$-operator content
 can read off directly  from $\jh$: all triplets of the form ($\jh_{i-1},\jh_{i},\jh_{i+1})=(\jh_{i}\mp1/2,\jh_{i},\jh_{i}\pm1/2)$
 indicates the action of $c_i$, while triplets of  the type $(\jh_{i}\pm1/2,\jh_{i},\jh_{i}\pm1/2)$ correspond to no operator action.\footnote{Alternatively, 
 a $c_i$ action is signaled by a  pair of consecutive operators in $\pi_b$ of rank $i$ and $i+1$ that do not satisfy the repulsion conditions (\ref{succ}).} Take for instance the reduced vector (\ref{jhex}) and indicate in bold the middle entry in triplets of the form $(\jh_{i}\mp1/2,\jh_{i},\jh_{i}\pm1/2)$:
 \begin{equation}\label{jhexa}
	\jh(\pi_b)= \l(0,{\bf \tfrac12},{\bf1},\tfrac32,{\bf1},\tfrac12,{\bf1},\tfrac32,1,\tfrac32,{\bf1},{\bf\tfrac12},0\r).
\end{equation}
From the positions of these bold entries, $1,2,4,6,10,11$, we directly infer $\pi_c$ to be $c_1c_2c_4c_6c_{10}c_{11}$.

With $\pi_c$ determined, it remains to get $\pi^0_b$. This amounts to undo the action of $\pi_c^{\text{rev}}$ on $\pi_b$, i.e., remove  the $\cip{i}$ successively, in decreasing value of the rank. Each removal transforms $\pi_b$ by modifying the positions of the operators, (i.e., the entries of $\vec l$), as
\begin{align}
	l_m\rw l_m+1\quad \forall m>i.
\end{align}
By construction, once the action of all elements of $\pi_c^{\text{rev}}$ have been undone, the remaining sequence is  $\pi_b^0$.\footnote{It should be clear that the resulting sequence is an element of $\Pi^{(1)}$, meaning that the repulsion conditions (\ref{succ}) are satisfied. This, as previously indicated, is equivalent to have all $j_l$ in the range $0\leq j_l\leq \frac12$, implying that $0\leq \jh_i\leq \frac12$.  But removing the action of $\pi_c$ on $\jh$ leaves $\jh^{(0)}$, ensuring thus that the resulting corresponding sequence is indeed a restricted one.}
This is how the $\pi_b^0$ sequence in the example of Section \ref{operateursci} (cf. Eq. (\ref{excsurba})) is obtained. Indeed, denoting by $c_i^{-1}$ the operation of undoing the action of $c_i$, we have:
\begin{align}\label{exlc}
	\vec l=&\,
(\underline{0},\underline{0},\underline{0},2,2,3,3,4,\underline{6},\underline{7},\underline{7},\underline{7}),
\nonumber\\
c_{11}^{-1}:&\,(\underline{0},\underline{0},\underline{0},2,2,3,3,4,\underline{6},\underline{7},\underline{7},\underline{8})
\nonumber\\
c_{10}^{-1}:&\,(\underline{0},\underline{0},\underline{0},2,2,3,3,4,\underline{6},\underline{7},\underline{8},\underline{9})
\nonumber\\
c_{6}^{-1}:&\,(\underline{0},\underline{0},\underline{0},2,2,3,4,5,\underline{7},\underline{8},\underline{9},\underline{10})
\nonumber\\
c_{4}^{-1}:&\,(\underline{0},\underline{0},\underline{0},2,3,4,5,6,\underline{8},\underline{9},\underline{10},\underline{11})
\nonumber\\
c_{2}^{-1}:&\,(\underline{0},\underline{0},\underline{1},3,4,5,6,7,\underline{9},\underline{10},\underline{11},\underline{12})
\nonumber\\
c_{1}^{-1}:&\,(\underline{0},\underline{1},\underline{2},4,5,6,7,8,\underline{10},\underline{11},\underline{12},\underline{13}),
\end{align}
which is the $\vec l$ vector associated to the restricted sequence $\pi_b^0$ of (\ref{excsurba}).

\section{Path interpretation of the reduced vector: $\ckp{k}$ paths}\label{chejk}

Let us restate what we have reached at this point and what is our immediate goal. The $\mkp{k}$ paths are in one-to-one correspondence with general sequences $\pi_b^{\text{rev}}$, acting on $\ga^{\gs(0;k)}$. 
All such sequences $\pi_b\in\Pi^{(k)}(n_+,n_-)$, 
 are generated by strings $\pi_c$ acting on restricted sequences $\pi_b^0\in\Pi^{(1)}(n_+,n_-)$.
  The $\pi_b^0$  sequences are in bijection with $k=1$ paths, so that their generating function is known. Our quest now is to find the generating function of the sequences $\pi_b$ generated on the top of $\pi_b^0$, namely, the generating function for sequences $\pi_c$.
The generating function of the $\mkp{k}$ paths with specified boundary condition, i.e., the $\su{k}$ character for the corresponding module, is the product of these two generating functions, summed over $n_\pm$.

We are now in position to present our main trick for the $q$-enumeration of all admissible sequences of $c_i$ on $\pi_b^0$. For this we recall that the action of $c_i$ on some $\pi_b$ can equally well be described by its action on the corresponding reduced vector $\jh$. 
As we will see in the following subsection, $\jh$ itself, and therefore the action of the operators $c_i$ on $\jh$, can be dressed by a path interpretation, a crucial step for the derivation  of the generating function.

\subsection{From reduced vectors to $\ckp{k}$ paths}

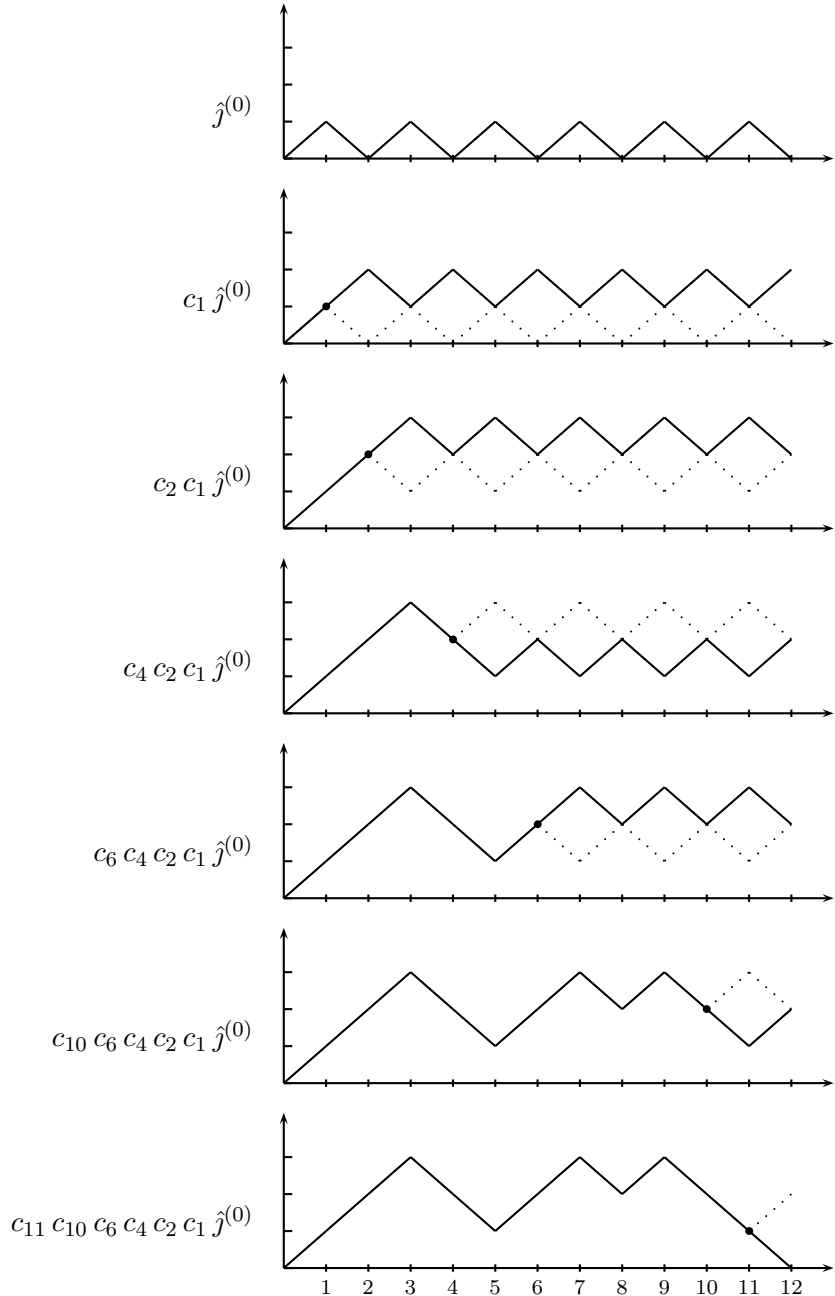
\begin{figure}[htpb]\
\caption{The step by step construction of the $\ckp{3}_0(12)$ path associated to the $\jh$ vector given in Eq. (\ref{jhex}), i.e., the path transposition of the construction worked out in Eq. (\ref{construction jh}). Units on the vertical axis are half-integers. The application point of the different $c_i$ are indicated by dots. The  portion of the path that has been flipped, upward or downward from the application point, is indicated by dotted lines. The weight $\wt$ of the final path calculated from (\ref{wdemu}) is $\wt=-\frac12(11+10+8+6+2+1)=-19$. Finally, the value of $j$ is given by vertical position of the final point, which 1 for all paths but the first and the last ones.}

\label{construction ci}
\begin{center}
\begin{pspicture}(2,0)(3,17)
{\psset{yunit=14pt,xunit=16pt,linewidth=.8pt}
\rput[Br](0.3,31){{ $\jh^{(0)}$}}
\rput[Br](0.3,26){{ $c_1\,\jh^{(0)}$}}
\rput[Br](0.3,21){{ $c_2\,c_1\,\jh^{(0)}$}}
\rput[Br](0.3,16){{ $c_4\,c_2\,c_1\,\jh^{(0)}$}}
\rput[Br](0.3,11){{ $c_6\,c_4\,c_2\,c_1\,\jh^{(0)}$}}
\rput[Br](0.3,6){{ $c_{10}\,c_6\,c_4\,c_2\,c_1\,\jh^{(0)}$}}
\rput[Br](0.3,1){{ $c_{11}\,c_{10}\,c_6\,c_4\,c_2\,c_1\,\jh^{(0)}$}}

\psline{->}(1,0)(1,4.2)
\psline{->}(1,5)(1,9.2)
\psline{->}(1,10)(1,14.2)
\psline{->}(1,15)(1,19.2)
\psline{->}(1,20)(1,24.2)
\psline{->}(1,25)(1,29.2)
\psline{->}(1,30)(1,34.2)
\psline{->}(1,0)(14,0)
\psline{->}(1,5)(14,5)
\psline{->}(1,10)(14,10)
\psline{->}(1,15)(14,15)
\psline{->}(1,20)(14,20)
\psline{->}(1,25)(14,25)
\psline{->}(1,30)(14,30)

\psline{-}(1,0)(1.2,0)\psline{-}(1,1)(1.2,1)\psline{-}(1,2)(1.2,2)\psline{-}(1,3)(1.2,3)
\psline{-}(1,5)(1.2,5)\psline{-}(1,6)(1.2,6)\psline{-}(1,7)(1.2,7)\psline{-}(1,8)(1.2,8)
\psline{-}(1,10)(1.2,10)\psline{-}(1,11)(1.2,11)\psline{-}(1,12)(1.2,12)\psline{-}(1,13)(1.2,13)
\psline{-}(1,15)(1.2,15)\psline{-}(1,16)(1.2,16)\psline{-}(1,17)(1.2,17)\psline{-}(1,18)(1.2,18)
\psline{-}(1,20)(1.2,20)\psline{-}(1,21)(1.2,21)\psline{-}(1,22)(1.2,22)\psline{-}(1,23)(1.2,23)
\psline{-}(1,25)(1.2,25)\psline{-}(1,26)(1.2,26)\psline{-}(1,27)(1.2,27)\psline{-}(1,28)(1.2,28)
\psline{-}(1,30)(1.2,30)\psline{-}(1,31)(1.2,31)\psline{-}(1,32)(1.2,32)\psline{-}(1,33)(1.2,33)

\psline{-}(2,-0.1)(2,0.1)\psline{-}(3,-0.1)(3,0.1)\psline{-}(4,-0.1)(4,0.1)\psline{-}(5,-0.1)(5,0.1)\psline{-}(6,-0.1)(6,0.1)
\psline{-}(7,-0.1)(7,0.1)\psline{-}(8,-0.1)(8,0.1)\psline{-}(9,-0.1)(9,0.1)\psline{-}(10,-0.1)(10,0.1)\psline{-}(11,-0.1)(11,0.1)
\psline{-}(12,-0.1)(12,0.1)\psline{-}(13,-0.1)(13,0.1)

\psline{-}(2,4.9)(2,5.1)\psline{-}(3,4.9)(3,5.1)\psline{-}(4,4.9)(4,5.1)\psline{-}(5,4.9)(5,5.1)\psline{-}(6,4.9)(6,5.1)
\psline{-}(7,4.9)(7,5.1)\psline{-}(8,4.9)(8,5.1)\psline{-}(9,4.9)(9,5.1)\psline{-}(10,4.9)(10,5.1)\psline{-}(11,4.9)(11,5.1)
\psline{-}(12,4.9)(12,5.1)\psline{-}(13,4.9)(13,5.1) 

\psline{-}(2,9.9)(2,10.1)\psline{-}(3,9.9)(3,10.1)\psline{-}(4,9.9)(4,10.1)\psline{-}(5,9.9)(5,10.1)\psline{-}(6,9.9)(6,10.1)
\psline{-}(7,9.9)(7,10.1)\psline{-}(8,9.9)(8,10.1)\psline{-}(9,9.9)(9,10.1)\psline{-}(10,9.9)(10,10.1)\psline{-}(11,9.9)(11,10.1)
\psline{-}(12,9.9)(12,10.1)\psline{-}(13,9.9)(13,10.1)

\psline{-}(2,14.9)(2,15.1)\psline{-}(3,14.9)(3,15.1)\psline{-}(4,14.9)(4,15.1)\psline{-}(5,14.9)(5,15.1)\psline{-}(6,14.9)(6,15.1)
\psline{-}(7,14.9)(7,15.1)\psline{-}(8,14.9)(8,15.1)\psline{-}(9,14.9)(9,15.1)\psline{-}(10,14.9)(10,15.1)\psline{-}(11,14.9)(11,15.1)
\psline{-}(12,14.9)(12,15.1)\psline{-}(13,14.9)(13,15.1) 

\psline{-}(2,19.9)(2,20.1)\psline{-}(3,19.9)(3,20.1)\psline{-}(4,19.9)(4,20.1)\psline{-}(5,19.9)(5,20.1)\psline{-}(6,19.9)(6,20.1)
\psline{-}(7,19.9)(7,20.1)\psline{-}(8,19.9)(8,20.1)\psline{-}(9,19.9)(9,20.1)\psline{-}(10,19.9)(10,20.1)\psline{-}(11,19.9)(11,20.1)
\psline{-}(12,19.9)(12,20.1)\psline{-}(13,19.9)(13,20.1) 

\psline{-}(2,24.9)(2,25.1)\psline{-}(3,24.9)(3,25.1)\psline{-}(4,24.9)(4,25.1)\psline{-}(5,24.9)(5,25.1)\psline{-}(6,24.9)(6,25.1)
\psline{-}(7,24.9)(7,25.1)\psline{-}(8,24.9)(8,25.1)\psline{-}(9,24.9)(9,25.1)\psline{-}(10,24.9)(10,25.1)\psline{-}(11,24.9)(11,25.1)
\psline{-}(12,24.9)(12,25.1)\psline{-}(13,24.9)(13,25.1) 

\psline{-}(2,29.9)(2,30.1)\psline{-}(3,29.9)(3,30.1)\psline{-}(4,29.9)(4,30.1)\psline{-}(5,29.9)(5,30.1)\psline{-}(6,29.9)(6,30.1)
\psline{-}(7,29.9)(7,30.1)\psline{-}(8,29.9)(8,30.1)\psline{-}(9,29.9)(9,30.1)\psline{-}(10,29.9)(10,30.1)\psline{-}(11,29.9)(11,30.1)
\psline{-}(12,29.9)(12,30.1)\psline{-}(13,29.9)(13,30.1) 

\psline{-}(1,0)(2,1)
\psline{-}(2,1)(3,2)
\psline{-}(3,2)(4,3)
\psline{-}(4,3)(5,2)
\psline{-}(5,2)(6,1)
\psline{-}(6,1)(7,2)
\psline{-}(7,2)(8,3)
\psline{-}(8,3)(9,2)
\psline{-}(9,2)(10,3)
\psline{-}(10,3)(11,2)
\psline{-}(11,2)(12,1)
\psline{-}(12,1)(13,0) 
 
\psline{-}(1,5)(2,6)
\psline{-}(2,6)(3,7)
\psline{-}(3,7)(4,8)
\psline{-}(4,8)(5,7)
\psline{-}(5,7)(6,6)
\psline{-}(6,6)(7,7)
\psline{-}(7,7)(8,8)
\psline{-}(8,8)(9,7)
\psline{-}(9,7)(10,8)
\psline{-}(10,8)(11,7)
\psline{-}(11,7)(12,6)
\psline{-}(12,6)(13,7)

\psline{-}(1,10)(2,11)
\psline{-}(2,11)(3,12)
\psline{-}(3,12)(4,13)
\psline{-}(4,13)(5,12)
\psline{-}(5,12)(6,11)
\psline{-}(6,11)(7,12)
\psline{-}(7,12)(8,13)
\psline{-}(8,13)(9,12)
\psline{-}(9,12)(10,13)
\psline{-}(10,13)(11,12)
\psline{-}(11,12)(12,13)
\psline{-}(12,13)(13,12)

\psline{-}(1,15)(2,16)
\psline{-}(2,16)(3,17)
\psline{-}(3,17)(4,18)
\psline{-}(4,18)(5,17)
\psline{-}(5,17)(6,16)
\psline{-}(6,16)(7,17)
\psline{-}(7,17)(8,16)
\psline{-}(8,16)(9,17)
\psline{-}(9,17)(10,16)
\psline{-}(10,16)(11,17)
\psline{-}(11,17)(12,16)
\psline{-}(12,16)(13,17)

\psline{-}(1,20)(2,21)
\psline{-}(2,21)(3,22)
\psline{-}(3,22)(4,23)
\psline{-}(4,23)(5,22)
\psline{-}(5,22)(6,23)
\psline{-}(6,23)(7,22)
\psline{-}(7,22)(8,23)
\psline{-}(8,23)(9,22)
\psline{-}(9,22)(10,23)
\psline{-}(10,23)(11,22)
\psline{-}(11,22)(12,23)
\psline{-}(12,23)(13,22)

\psline{-}(1,25)(2,26)
\psline{-}(2,26)(3,27)
\psline{-}(3,27)(4,26)
\psline{-}(4,26)(5,27)
\psline{-}(5,27)(6,26)
\psline{-}(6,26)(7,27)
\psline{-}(7,27)(8,26)
\psline{-}(8,26)(9,27)
\psline{-}(9,27)(10,26)
\psline{-}(10,26)(11,27)
\psline{-}(11,27)(12,26)
\psline{-}(12,26)(13,27)

\psline{-}(1,30)(2,31)
\psline{-}(2,31)(3,30)
\psline{-}(3,30)(4,31)
\psline{-}(4,31)(5,30)
\psline{-}(5,30)(6,31)
\psline{-}(6,31)(7,30)
\psline{-}(7,30)(8,31)
\psline{-}(8,31)(9,30)
\psline{-}(9,30)(10,31)
\psline{-}(10,31)(11,30)
\psline{-}(11,30)(12,31)
\psline{-}(12,31)(13,30)

\psset{linestyle=dotted}
\psline{-}(12,1)(13,2)
\psline{-}(11,7)(12,8)
\psline{-}(12,8)(13,7)
\psline{-}(7,12)(8,11)
\psline{-}(8,11)(9,12)
\psline{-}(9,12)(10,11)
\psline{-}(10,11)(11,12)
\psline{-}(11,12)(12,11)
\psline{-}(12,11)(13,12)
\psline{-}(5,17)(6,18)
\psline{-}(6,18)(7,17)
\psline{-}(7,17)(8,18)
\psline{-}(8,18)(9,17)
\psline{-}(9,17)(10,18)
\psline{-}(10,18)(11,17)
\psline{-}(11,17)(12,18)
\psline{-}(12,18)(13,17)
\psline{-}(3,22)(4,21)
\psline{-}(4,21)(5,22)
\psline{-}(5,22)(6,21)
\psline{-}(6,21)(7,22)
\psline{-}(7,22)(8,21)
\psline{-}(8,21)(9,22)
\psline{-}(9,22)(10,21)
\psline{-}(10,21)(11,22)
\psline{-}(11,22)(12,21)
\psline{-}(12,21)(13,22)
 \psline{-}(1,25)(2,26)
\psline{-}(2,26)(3,25)
\psline{-}(3,25)(4,26)
\psline{-}(4,26)(5,25)
\psline{-}(5,25)(6,26)
\psline{-}(6,26)(7,25)
\psline{-}(7,25)(8,26)
\psline{-}(8,26)(9,25)
\psline{-}(9,25)(10,26)
\psline{-}(10,26)(11,25)
\psline{-}(11,25)(12,26)
\psline{-}(12,26)(13,25)

\rput(2,-0.5){{\scriptsize $1$}}
\rput(3,-0.5){{\scriptsize $2$}}
\rput(4,-0.5){{\scriptsize $3$}}
\rput(5,-0.5){{\scriptsize $4$}}
\rput(6,-0.5){{\scriptsize $5$}}
\rput(7,-0.5){{\scriptsize $6$}}
\rput(8,-0.5){{\scriptsize $7$}}
\rput(9,-0.5){{\scriptsize $8$}}
\rput(10,-0.5){{\scriptsize $9$}}
\rput(11,-0.5){{\scriptsize $10$}}
\rput(12,-0.5){{\scriptsize $11$}}
\rput(13,-0.5){{\scriptsize $12$}}

\psset{dotsize=3pt}\psset{dotstyle=*}
\psdots(2.0,26.0)(3,22)(5,17)(7,12)(11,7)(12,1)
}
\end{pspicture}
\end{center}
\end{figure}

To make manifest the (finite) path interpretation of the vector $\jh$, set $ N=n_++n_- $ and rewrite the vector as 
\begin{align}\jh=(\jh_0,\jh_1,\cdots, \jh_{N})\quad\text{where}\quad \jh_0=0,\;\jh_N=j,\;\text{and}\;|\jh_{i}-\jh_{i+1}|=\frac12.
\end{align}
By considering the entries of this vector as vertices that are joined by edges, one gets a path, also denoted by $\jh$.
The set of such paths is  denoted $\ckp{k}_j(N)$, examples of which are presented in Fig. \ref{construction ci}. 
By re-scaling $\jh\to p_i=2\jh_i$, a $\ckp{k}$ path looks very much like a RSOS path (with height in the range $0\leq p_i\leq k$).

This relationship between $\ckp{k}$ and RSOS paths goes a step further: as we now argued, the weight of a $\ckp{k}$-path is given by the RSOS-like expression:
\begin{align}\label{wdemu}
	\wc(\jh)=\sum_{i=1}^N\wc_i\quad\text{where}\quad \wc_i=\frac{1}{2}(i-N)|\jh_{i+1}-\jh_{i-1}|,
\end{align}
i.e., the weight of a straight vertex at $i$ (i,.e., of the form $(\jh_{i}\mp1/2,\jh_{i},\jh_{i}\pm1/2)$) is $(i-N)/2$, while the peaks and valleys have weight zero. To see whether this weight has to be evaluated with respect to some ground state with fixed boundaries,  recall that the goal is to $q$-enumerate the vectors $\jh$ (for all $0\leq j\leq k/2$) built on $\jh^{(0)}$ by sequences of $c_i$; the weight is thus relative to that of $\jh^{(0)}$.  But this brings no correction since $\wc(\jh^{(0)})=0$.

Now, the action of $c_i$ on $\jh$ is to create a straight vertex (up or down) at the position of a local extremum (peak or valley respectively). It thus follows from (\ref{wdemu}) that
\begin{align}\label{wdec}
	\wc(c_i\jh)-\wc(\jh)=\frac{1}{2}(i-N).
\end{align}
The comparison of this expression with (\ref{wec}) (recalling that $ N=n_++n_- $) proves (\ref{wdemu}).

Pictorially, the action of $c_i$ is to flip the oscillating tail of the path for $n\geq i$ with respect to the horizontal line at height $\jh_i$, upward if $\jh_i$ is a peak or downward if $\jh_i$ is a valley. The portion of the path that have been flipped at each step in Fig. \ref{construction ci}  is indicated by dotted lines.

$\ckp{k}$ paths are thus analogous to the RSOS paths describing the finitization of the states in irreducible modules of type $(r,s)$, with $r=2\jh_0+1=1$ and $s=2j_N+1=2j+1$, for the minimal model $\mathcal M(k+1,k+2)$. The only difference is that the weight is slightly modified by the $-N$ term. Moreover, the $c_i$ operator description is the direct transposition of that in \cite{JM}.\footnote{The operators are denoted as $b_i$ and $b_i^*$ in \cite{JM}: the operator $c_i$ acts as a $b_i$ on a peak  and a $b_i^*$ on a valley. They are  also equivalent to $\ab^{\pm1}$ operators  introduced earlier in the case $k=1$, in the context of $\akp{1}$ paths \cite{LPM}.} 

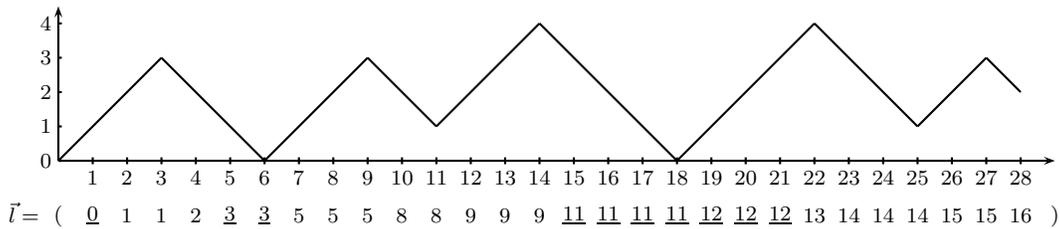
\begin{figure}[ht]\
\caption{The $\ckp{4}_1(28)$  path associated to the $\mk^{(4)}_1$ path of Fig. \ref{exS4}, drawn with the entries of $\jh$ multiplied by 2 (i.e., as a RSOS path). $\jh$ 
is constructed from the vector $\vec l$ indicated below, itself read from the vector $\vec a$ given in Fig. \ref{exS4}. The relation between $\vec l$ and the pictured path is the following: the height at position $i$ is that at $i-1$ plus  ${\e_i}(-1)^{l_i-1}$ where $\e_i=-1$ if $l_i$ is underlined and 1 otherwise (cf. Eq. (\ref{jil})). }
\label{rsos4} 
\begin{center}
\begin{pspicture}(4.5,-1)(8,1.9)
{\psset{yunit=13pt,xunit=13pt,linewidth=.8pt}
\psline{->}(0,0.)(0,4.5) \psline{->}(0,0)(29,0)
\rput(2,-0.5){{\scriptsize $2$}}
\rput(1,-0.5){{\scriptsize $1$}}
\rput(3,-0.5){{\scriptsize $3$}}
\rput(5,-0.5){{\scriptsize $5$}}
\rput(7,-0.5){{\scriptsize $7$}}
\rput(9,-0.5){{\scriptsize $9$}}
\rput(11,-0.5){{\scriptsize $11$}}
\rput(13,-0.5){{\scriptsize $13$}}
\rput(4,-0.5){{\scriptsize $4$}}
\rput(6,-0.5){{\scriptsize $6$}}
\rput(8,-0.5){{\scriptsize $8$}}
\rput(10,-0.5){{\scriptsize $10$}}
\rput(12,-0.5){{\scriptsize $12$}}
\rput(14,-0.5){{\scriptsize $14$}}
\rput(15,-0.5){{\scriptsize $15$}}
\rput(16,-0.5){{\scriptsize $16$}}
\rput(17,-0.5){{\scriptsize $17$}}
\rput(18,-0.5){{\scriptsize $18$}}
\rput(19,-0.5){{\scriptsize $19$}}
\rput(20,-0.5){{\scriptsize $20$}}
\rput(21,-0.5){{\scriptsize $21$}}
\rput(22,-0.5){{\scriptsize $22$}}
\rput(23,-0.5){{\scriptsize $23$}}
\rput(24,-0.5){{\scriptsize $24$}}
\rput(25,-0.5){{\scriptsize $25$}}
\rput(26,-0.5){{\scriptsize $26$}}
\rput(27,-0.5){{\scriptsize $27$}}
\rput(28,-0.5){{\scriptsize $28$}}

\rput[r](-0.2,2){{\scriptsize $2$}}
\rput[r](-0.2,0){{\scriptsize $0$}}
\rput[r](-0.2,1){{\scriptsize $1$}}
\rput[r](-0.2,3){{\scriptsize $3$}}
\rput[r](-0.2,4){{\scriptsize $4$}}

\psset{linestyle=solid}
\psline{-}(1,-0.1)(1,0.1)\psline{-}(2,-0.1)(2,0.1)
\psline{-}(3,-0.1)(3,0.1)\psline{-}(4,-0.1)(4,0.1)
\psline{-}(5,-0.1)(5,0.1)\psline{-}(6,-0.1)(6,0.1)
\psline{-}(7,-0.1)(7,0.1)\psline{-}(8,-0.1)(8,0.1)
\psline{-}(9,-0.1)(9,0.1)\psline{-}(10,-0.1)(10,0.1)
\psline{-}(11,-0.1)(11,0.1)\psline{-}(12,-0.1)(12,0.1)
\psline{-}(13,-0.1)(13,0.1)\psline{-}(14,-0.1)(14,0.1)
\psline{-}(15,-0.1)(15,0.1)\psline{-}(16,-0.1)(16,0.1)
\psline{-}(17,-0.1)(17,0.1)
\psline{-}(18,-0.1)(18,0.1)
\psline{-}(19,-0.1)(19,0.1)
\psline{-}(20,-0.1)(20,0.1)
\psline{-}(21,-0.1)(21,0.1)
\psline{-}(22,-0.1)(22,0.1)
\psline{-}(23,-0.1)(23,0.1)
\psline{-}(24,-0.1)(24,0.1)
\psline{-}(25,-0.1)(25,0.1)
\psline{-}(26,-0.1)(26,0.1)
\psline{-}(27,-0.1)(27,0.1)
\psline{-}(28,-0.1)(28,0.1)

\psline{-}(0,1)(.1,1)
\psline{-}(0,2)(.1,2)
\psline{-}(0,3)(.1,3)
\psline{-}(0,0)(.1,0)
\psline{-}(0,4)(.1,4)

\psline{-}(0,0)(3,3)
\psline{-}(3,3)(6,0)
\psline{-}(6,0)(9,3)
\psline{-}(9,3)(11,1)
\psline{-}(11,1)(14,4)
\psline{-}(14,4)(18,0)
\psline{-}(18,0)(22,4)
\psline{-}(22,4)(25,1)
\psline{-}(25,1)(27,3)\psline{-}(27,3)(28,2)

\rput(-1,-1.5){{\scriptsize $\vec l=$}}
\rput(0,-1.6){{\scriptsize $($}}
\rput(1,-1.6){{\scriptsize $\underline0$}}
\rput(2,-1.6){{\scriptsize $1$}}
\rput(3,-1.6){{\scriptsize $1$}}
\rput(4,-1.6){{\scriptsize $2$}}
\rput(5,-1.6){{\scriptsize $\underline3$}}
\rput(6,-1.6){{\scriptsize $\underline3$}}
\rput(7,-1.6){{\scriptsize $5$}}
\rput(8,-1.6){{\scriptsize $5$}}
\rput(9,-1.6){{\scriptsize $5$}}
\rput(10,-1.6){{\scriptsize $8$}}
\rput(11,-1.6){{\scriptsize $8$}}
\rput(12,-1.6){{\scriptsize $9$}}
\rput(13,-1.6){{\scriptsize $9$}}
\rput(14,-1.6){{\scriptsize $9$}}
\rput(15,-1.6){{\scriptsize $\underline{11}$}}
\rput(16,-1.6){{\scriptsize $\underline{11}$}}
\rput(17,-1.6){{\scriptsize $\underline{11}$}}
\rput(18,-1.6){{\scriptsize $\underline{11}$}}
\rput(19,-1.6){{\scriptsize $\underline{12}$}}
\rput(20,-1.6){{\scriptsize $\underline{12}$}}
\rput(21,-1.6){{\scriptsize $\underline{12}$}}
\rput(22,-1.6){{\scriptsize $13$}}
\rput(23,-1.6){{\scriptsize $14$}}
\rput(24,-1.6){{\scriptsize $14$}}
\rput(25,-1.6){{\scriptsize $14$}}
\rput(26,-1.6){{\scriptsize $15$}}
\rput(27,-1.6){{\scriptsize $15$}}
\rput(28,-1.6){{\scriptsize $16$}}
\rput(29,-1.6){{\scriptsize $)$}}

}
\end{pspicture}
\end{center}
\end{figure}

Summing up,  a path $\ga\in\mkp{k}_j(n_+,n_-)$ is represented by a pair of paths:
\begin{equation}
\ga\lrw (\ga',\jh)\quad \text{with}\quad\begin{cases}
& \ga'\in\mkp{1}_{j'}(n_+,n_-)\quad\text{with}\quad j'=\frac12[n_++n_-]_2,\\
&\jh\in\ckp{k}_j(n_++n_-).\end{cases}
\end{equation}
The $\ckp{3}_0$
 path associated to the $\mk^{(3)}_0$ path of Fig. \ref{exemplemk} is thus the last path in Fig. \ref{construction ci}. Similarly, the $\ckp{4}_1$ path corresponding to the $\mk^{(4)}_1$ path of Fig. \ref{exS4} is pictured (with rescaled vertical coordinates) in Fig. \ref{rsos4}.

\subsection{Generating function for $\ckp{k}$ paths}\label{generc}

Let $\Phi_{j;k}(N;q)$ be the generating function of $\ckp{k}_j(N)$ paths: 
\begin{align}
	\Phi_{j;k}(N;q)=\sum_{\jh\in\ckp{k}_j(N)}q^{\wc(\jh)}.
\end{align}
To find this generating function, we use the fact that it is equal, except for the $-N$ term in the weight, to the well-known finitized form of the $\mathcal M(k+1,k+2)$ 
characters \cite{FW,OleJSb}.  But instead of inserting directly the $-N$ correction, we introduce the reversed path 
$\jh^{\text{rev}}=(\jh^{\text{rev}}_0 ,\jh^{\text{rev}}_1,\cdots, \jh^{\text{rev}}_{N})$ 
defined as follow:
\begin{align}
	\jh_{i}^{\text{rev}}=\jh_{N-i}.
\end{align}
The interest of this path reversal is that the weight becomes
\begin{align}
	\wcp{i}^{\text{rev}}&=\wcp{N-i}= 
-\frac12i|\jh_{i+1}^{\text{rev}}-\jh_{i-1}^{\text{rev}}|,
\end{align}
which is precisely minus the standard RSOS weight (with the standard RSOS height given by $p_i=2\jh_i$) \cite{ABF}.

Thus the sought-for generating function $\Phi_{j;k}(N;q)$ is dual (i.e., obtained by the transformation $q\rw q^{-1}$) to that of the finitized $\mathcal M(k+1,k+2)$-type paths 
(the precise match would follow by the replacement $\jh_i\to p_i=2\jh_i$) and with boundary conditions $\jh_{0}^{\text{rev}}=\jh_{N}=j$ and $\jh_{N}^{\text{rev}}=\jh{0}=0$. Denote the latter generating function by $\tilde{\Phi}_{j;k}(N;q)$. There are actually two known  fermionic  forms $	\tilde{\Phi}^{[\ell]}_{j;k }(N;q)$, $\ell=1,2$, for this generating function, given in \cite{Mel2}:\footnote{See also \cite{FW}, Theorem 5.1, the third and fourth equations (pertaining to the cases $b=1$) or \cite[Eqs. (2.1) and (2.4)]{OleJSb}.}
\begin{align}\label{phim1}
	\tilde{\Phi}^{[\ell]}_{j;k }(N;q)=\sum_{\substack{m_1,...,m_{k-1}\geq  0 \\m_i\equiv Q^{[\ell]}_i (\text{mod }2)}} 
	\phi^{[\ell]}_{t_\ell;k }(\{m_i\};m_{t_\ell};N;q)
		\end{align}
	where
\begin{align}\label{phim2} 		\phi^{[\ell]}_{t_\ell;k }(\{m_i\};m_{t_\ell};N;q)=  	q^{\frac12\sum_{i=1}^{k-1}(m_i^2-m_im_{i-1})-\frac12m_{t_\ell}}\prod_{i=1}^{k-1}\qbin{\frac12(m_{i-1}+m_{i+1}+N\delta_{i,1}+\delta_{i,t_\ell})}{m_i},
\end{align}
with  $m_0=m_k=0$ and
\begin{align}\label{phim3}
t_1=k-2j, \qquad t_2=2j,  \qquad  Q^{[1]}_i=\min(2j,k-i),\qquad Q^{[2]}_i=2j
+\min(2j,i).
\end{align}
(To obtain these expressions for $Q^{[\ell]}$ from those given in \cite{Mel2,FW,OleJSb}, we use the identity (3.35) of the latter reference.)
The $q$-binomial is defined as
\begin{equation}
\begin{bmatrix}
a\\ b\end{bmatrix} \equiv \begin{bmatrix}
a\\ b\end{bmatrix}_q =  \begin{cases} & \frac{(q)_a}{(q)_{a-b}(q)_b}  \quad \text{if}\quad 0\leq b\leq a,\\ &\qquad 0\qquad\qquad\text{otherwise}.
\end{cases}
\end{equation}
 The dual character is obtained by the inversion $q\rw q^{-1}$, 
that is
\begin{align}
	\Phi_{j;k}^{[\ell]}(N;q)= \tilde{\Phi}^{[\ell]}_{j;k }(N; q^{-1}).
	\end{align}
 Using the easily derived identity 
\begin{align}
	\qbin{n}{m}_{q^{-1}}=q^{m(m-n)}\qbin{n}{m}_{q},
\end{align}
and simple  algebraic  manipulations, one finds:
\begin{align}\label{phi1}
	\Phi^{[\ell]}_{j;k}(N;q)=\sum_{\substack{m_1,...,m_{k-1}\geq 0\\m_i\equiv Q_i^{[\ell]}(\text{mod }2)}}
				\phi^{[\ell]}_{t_\ell;k }(\{m_i\};Nm_{1};N;q).
\end{align}
This is thus the generating function 
for all sequences $\pi_c$ acting on $\jh^{(0)}=(0,\frac12,\cdots,\frac12[N]_2)$.
As already mentioned, this expression is for a fixed sector and this is the reason for the absence of a $z$ dependence.

\subsection{The particle content of RSOS paths}

Our claim to provide a constructive proof for the $\su{k}$ fermionic characters felt short at the last step where we rather invoke known results. But the first form $\Phi^{[1]}_{j;k}(N;q)$ is actually obtained constructively in \cite{OleJSb} and there is no point in repeating this analysis. The RSOS particles are clearly identified there (see also \cite{BreL} and \cite[Section 6]{Siena} for an alternative presentation in a somewhat different context). These are the elementary triangles whose height, when the two base points of the triangle lie on the axis (like the left-most triangle in Fig. \ref{rsos4}), is called the charge. Within a charge complex,
that is, within portions of the paths delimited by two points on the horizontal axis 
(one of which could be replaced by an extremity point) and containing more than one peak (for example, between the position 8 and 18 in Fig. \ref{rsos4}), it requires a criterion more  precise that  the mere height and it is formulated as follows. For the highest peak of the complex, the charge is again the height. If two peaks in a complex have the same height, 
it is the leftmost peak which is attributed the largest charge. 
For the other peaks in the complex, we proceed as follows: start from a peak and go down in both directions following the path profile; from the first point (that closest to the peak) at which the path changes its direction or stops, on one side or the other, draw a baseline from which the  height is read: this height is the charge.
This rule holds for a standard RSOS path, or its $q\to q^{-1}$ dual (e.g., see \cite{PMjmp}).
For the example of Fig. \ref{rsos4}, the peaks have charge (from left to right): $3,2,4,4,1$. 
The particle spectrum of a RSOS path with integer heights between 0 and $k$ corresponds thus to particles of charge $j$ ranging from  1 to $k$.

In the language of \cite{FLPW}, the difference between two heights in the defining strip within which RSOS paths are defined is called a band. Our level-$k$ RSOS  strip ($0\leq 2\jh_i\leq k$) has thus $k+1$ bands. In the unitary case (which is our situation), all bands are odd (or dark). Each dark band corresponds to a RSOS vacuum. Therefore a charge-1 particle does not interpolate between different vacua. But this is so for all particles of charge $t>1$. They are thus like a pair of kink-antikink of topological charge $t-1$. In our construction, they  appear as $su(2)$ singlets (there is no $z$ dependence in the RSOS piece).

It should be stressed that the label counting the particles of charge $i$ is not the $m_i$ introduced in (\ref{phim1})-(\ref{phim2}) but rather the $n_i$ defined such that the upper term in the $q$-binomial of (\ref{phim2}) reads $m_i+n_i$, i.e., \cite{Ber,BMlmp,OleJS,FW}
\begin{equation}
n_i=\frac12(m_{i-1}+m_{i+1}-2m_i+N\delta_{i,1}+\delta_{i,k-2j})
\end{equation}
In the operator language, an isolated  particle of charge $i$ is described by: 
\begin{equation}
c_n\,c_{n+1}\cdots c_{n+i-2}\,c_{n+i} \,c_{n+i+1}\cdots c_{n+2i-2},
\end{equation}
i.e., a sequence of $2(i-1)$ closely packed operators except for a gap of two in the middle.

The other formula (for $\ell=2)$ -- the one that makes contact with the spinon character of \cite{BLSb} -- is demonstrated by a combinatorial-type recurrence method in \cite{FW}. For this second form, we have also been able to find a constructive derivation, whose presentation will however be omitted.

\section{The $\su{k}$ fermionic character}\label{fermica}

Having generated every $\mkp{k}$ path in terms of a level-1 path and a $\ckp{k}$ one, we can now write their generating function.
With fixed operator content, it  is given by the product of the generating function $G_0(n_+,n_-;z;q)$ in (\ref{chark1d}), counting the elements of  $\Pi^{(1)}(n_+,n_-)$, times  
the generating function $\Phi^{[\ell]}_{j;k}(n_+,n_-;q)$ in (\ref{phim1})--(\ref{phim3}), counting $\ckp{k}_j(N)$ paths, and which comes in two versions. It only remains to relate $2j$ to the parity of the total number operator. Since $s+j\in\mathbb{Z}$ and that $2s=n_+-n_-$, it follows that $n_-+n_+\equiv 2j\, (\text{mod }2)$.
Note that so far both $\pi^0_b$ sequences and $\jh$ paths have 
 been weighted by $\wt$. The actual weight being $\wt-j/2$, we add the $-j/2$ correction term and sum the result over $n_\pm$:
 \begin{align}\label{carak}
	\chi_{j;k}^{[\ell]}(z;q)=\sum_{\substack{n_-,n_+\geq0\\n_-+n_+\equiv 2j\,(\text{mod }2)}}
		\frac{q^{\frac{1}{4}(n_-+n_+)^2-\frac{1}{2}j}z^{\frac{1}{2}(n_+-n_-)}}{(q)_{n_-}(q)_{n_+}}	\Phi^{[\ell]}_{j;k}(n_++n_-;q).
\end{align}
Alternatively, we can express this formula in terms of the total operator number  $N=n_-+n_+$ and the rescaled sector $S=2\ca=(n_+-n_-)$
\begin{align}
	\chi_{j;k}^{[\ell]}(z;q)=\sum_{N\equiv 2j\,(\text{mod }2)}\,\sum_{S=-N}^N
		\frac{q^{\frac{1}{4}N^2-\frac{1}{2}j}z^{\frac{1}{2}S}}{(q)_{\frac{1}{2}(N+S)}(q)_{\frac{1}{2}(N-S)}}\Phi^{[\ell]}_{j;k}(N;q).
\end{align}
For $\ell=2$, this is precisely the (normalized) level-$k$ 
spinon-type character formula \cite{BLSb}\footnote{The labels are slightly different in \cite{BLSb}; to recover their expression,  we must replace  $N\rw m_1$ $S\rw2j'$ and $m_i\rw m_{i+1}$ for $i=1,2,...,k-1$. }
where, using (\ref{phim1})--(\ref{phim3}),
\begin{align}
\Phi_{j;k}^{[2]} (N;q) 
	&	=\sum_{\substack{m_1,...,m_{k-1}\geq 0\\m_i\equiv2j+\text{min}(2j,i)\,(\text{mod }2)}}	q^{\frac12\sum_{i=1}^{k-1}(m_i^2-m_im_{i-1})-\frac12Nm_1}\prod_{i=1}^{k-1}
		\qbin{\frac12(m_{i-1}+m_{i+1}+N\delta_{i,1}+\delta_{i,2j})}{m_i}.
\end{align}
The summation condition implies that $m_{2j-1}, m_{2j-3}\cdots$ are odd and all the other $m_i$ are even. The form $\ell=1$ differs only in the replacement $\delta_{i,2j}\to\delta_{i,k-2j}$ and the parity condition  on $m_i$ which is changed to $m_i\equiv \text{min}(2j,k-i)\,(\text{mod }2)$. Clearly, both expressions are equivalent when $j=0$ and $k/2$.
 
For instance, for the case $k=2$ and the variant $\ell=2$, this formula reduces to
\begin{align}\label{cas2}
	\chi_{j;2}^{[2]}(z;q)=\sum_{\substack{n_-,n_+,m_1\geq0\\n_-+n_+\equiv2j\,(\text{mod }2)\\m_1\equiv 2j+\text{min}(2j,1)\,(\text{mod }2)}}
		\frac{q^{\frac{1}{4}(n_-+n_+)^2+\frac12(m_1^2-m_1(n_-+n_+))-\frac{1}{2}j}z^{\frac{1}{2}(n_+-n_-)}}{(q)_{n_-}(q)_{n_+}}
		\qbin{\frac12(n_-+n_++\delta_{j,\frac12})}{m_1}.
\end{align}
The sum on $m_1$ is thus even for $j=0,1/2$ and odd for $j=1$. For $\ell=1$, the only difference is that the restriction on $m_1$ is now $m_1\equiv \text{min}(2j,1)\,(\text{mod }2)$, which only modifies the $j=1/2$ case, where now $m_1$ is odd. 
These expressions differ from those derived in \cite{LPM}.
Finally, for $k=1$, the spinon formula of \cite{BPS,BLS} is recovered.

\section{Conclusion}
  
We  have thus presented a detailed analysis of the path representation of the states in integrable modules of the affine $\su{k}$ algebra \cite{DJKMO3,JMMO}, using from the start the new RSOS-like weighting (\ref{weightak})  of these paths.\footnote{
In that regard, note that the recovery of RSOS paths from tensor products of $\mkp{k}$ ones in \cite{JMMO} become much more natural in the light of this new weighting.}
At levels $k=1,2$, the analysis could be done as easily from the operator or the particle point of view \cite{LPM}.
However, the most convenient tool to investigate the higher-level paths is clearly the operator construction rather than the particle decomposition. On the one hand, the particles are not manifest at once from a higher-level path. On the other hand,  the extension  of the nonlocal operators representation of the $\mkp{1}$ paths is immediately generalized to arbitrary level. But nevertheless, the direct extension of  the method used at level 1 for constructing the generating function proves to be rather difficult.

To circumvent this difficulty, we have shown that the string of nonlocal operators (of $b$-type) acting on the vacuum ground-state path can be generated by a novel type of operators (the $c$ ones) acting on restricted sequences of $b$ operators, which sequences are equivalent to level-1 sequences and are thus in one-to-one correspondence with the associated level-1 paths.
These actions of $c$-operators on restricted strings of $b$-operators have next been interpreted in terms of paths. The latter paths are finitized RSOS paths representing the finitized states  in  unitary minimal models $\M(k+1,k+2)$.
The construction of the generating function for $\mkp{k}$ paths is then reduced to taking the product of two known factors: that for $\mkp{1}$ ones, with fixed operator content $(n_+,n_-)$, times the generating function for RSOS paths (of the  $\M(k+1,k+2)$-type), of length $n_++n_-$, the result being summed over $n_+,n_-$. For  the last step, we could make use the existing results \cite{OleJSb,FW}. This ends up in a new derivation of the $\su{k}$ fermionic character formula, where the observed factorization \cite{BLSb} into a $k=1$ spinon part and a RSOS one  is seen to be a reflection  of the path construction. And because the generating function for finite RSOS paths comes in two versions  \cite{Mel2,FW}, our approach yields, as a bonus, an alternative form of the usual spinon character -- an expression that appears to be new.

The representation of a level-$k$ path by a pair of paths, one at level-1 and the other being of RSOS type, has an unexpected offshoot since these two components have a clear path-particle decomposition. The resulting path particles are thus the particles at level 1 (which are spin-$\frac12$ doublets, the path realization of the kinks of \cite{FT}, interpolating between the modules $j=0$ and $j=\frac12$) and the usual ($su(2)$ singlet) RSOS kinks and anti-kinks (and no breathers). Such a description is also implicit in \cite{Ara}.

The nonlocal operators approach appears to be rather directly extendable to the $\widehat{su}(N)_k$ case. Actually, these operators turn out to have a natural interpretation in terms of the tableaux of \cite{DJKMO3,JMMO,FLOTW}, a point that we intend to discuss elsewhere.

\begin{appendix}


\section{Comparison with the Nagoya spectral decomposition}\label{Nag}

The authors of \cite{Ara} have obtained the decomposition of a set of $\mkp{k}$ paths sharing the same local energies into a RSOS path plus a Yangian piece captured by a Young diagram. This looks rather similar to our results, where a 
$\mkp{k}$ path is characterized by a pair $(\pi_c,\pi_b^0)$, which implies that  a number of $\mkp{k}$ paths share the
same RSOS path specified by $\pi_c$. 
The aim of this section is to show that the  resulting RSOS paths are actually the same in the two constructions and to sketch the relation between them.

In \cite{Ara}, the starting point 
is the construction, for a given path $\ga$, of the vector $\vec h=(h_0,h_1,h_2,\cdots)$, where $h_l$ is the local `energy' given by:\footnote{The point $h_0$ is not included in \cite{Ara}.}
\begin{equation}
h_l=H(\ga_l,\ga_{l+1})\qquad \text{where}\qquad H(\ga_l,\ga_{l+1}) = \frac12(k-\text{min}(\ga_l,-\ga_{l+1})),
\end{equation}
that it, $\wm^\circ(\ga)=\sum_l lh_l$.
With $\ga_0=-k$, it follows that $h_0=k$.  
The local energy  function $H$ is conveniently expressed in matrix form; for instance, for $k=3$, it reads
\begin{align}\label{H3}H(\ga_l,\ga_{l+1})=H_{\ga_l,\ga_{l+1}}=\begin{pmatrix}3&2&1&0\\ 3&2&1&1\\ 3&2&2&2\\ 3&3&3&3\end{pmatrix},
\end{align}
where rows and columns are ordered as $(3,1,-1,-3)$ (i.e., $H(3,-3)=0$).
 For $\ga\in\mkp{k}_j$, the tail of $\vec h$ matches that of 
 \begin{equation}
 \vec h^{\gs(j;k)}=(k-2j, 2j,k-2j,2j,k-2j,\cdots).
 \end{equation}
  This map $\ga\to \vec h$ is many to one.
As an example, (cf. \cite[Ex. 4.4]{Ara}), to the energy vector
\begin{align}
\vec h=(3,1,2,2,2,1,2,1,2,\cdots),
\label{h4.4}\end{align}
 with $k=3$ and $ j=\frac12$,
 there corresponds the six paths $\ga$ listed in Table \ref{tab1}.

\begin{table}[ht]
\caption{The six paths $\ga\in\mkp{3}$ with $j=1/2$ associated to the energy vector given in (\ref{h4.4}). The paths are all of the form $\ga=(-3,\ga_1,\ga_2,\ga_3,\ga_4,1,-1,1,-1,\cdots)$ with $(\ga_1,\ga_2,\ga_3,\ga_4)$ given by the entries of the first row.
 Adding the vertices $\ga_0=-3$ and  $\ga_5=1$ to each 4-tuple of this first row, we easily get the corresponding vector $\vec a=(a_0,a_1,a_2,a_3,a_4,0,0,\cdots)$  using (\ref{gavsa}), written as  $(a_0,a_1,a_2,a_3,a_4)$ in the second row. 
To these correspond the sequences $\pi_b$ listed below, from which the sector is read as $s=\frac12(n_+-n_-)$.
All these paths have the same weight $w=3$ and the same total number of operators $n_++n_-=3$. 
The different $\vec l=(l_1,l_2,l_3)$ are listed in the 5th row.
Using (\ref{jil}), all these vectors are seen to
correspond to the same $\jh$.
Note that all six sequences $\pi_b$ have their $|a_l|\leq 1$ but, since $\jh_2=1$, they are clearly not elements of $\Pi^{(1)}$ (which would require all $\jh_i\leq \frac12$). Actually, the straight-up vertex at position 1 in $\jh$ signals the action of $c_1$. Equivalently, the conditions (\ref{succ}) characterizing the restricted sequences $\pi^0_b$ are violated for each $\pi_b$ above, all between the first two operators, signaling again the action of $c_1$.
The different $\pi_b^0$  listed in the 7th row are obtained from $\pi_b$ by undoing the action of $c_1$, meaning increasing by 1 the application position of the second and third operator of $\pi_b$.}
\label{tab1}
\footnotesize{\begin{center}
\begin{tabular}{|c|c|c|c|c|c|c|} 
\hline
\hline
&&&&&&\\
$\ga$ & $(3,-1,1,1)$ & $(3,-1,-1,1)$ & $(1,-1,1,1)$ & $(3,-1,-1,-1)$ & $(1,-1,-1,1)$ & $(1,-1,-1,-1)$\\&&&&&&\\
$\vec a$&$(0,1,0,1,1)$&$(0,1,-1,0,1)$&$(-1,0,0,1,1)$&$(0,1,-1,-1,0)$&$(-1,0,-1,0,1)$&$(-1,0,-1,-1,0)$\\
&&&&&&\\ 
$\pi_b$&$ b_1b_3b_4$&$ b_1b_2^{-1}b_4$&$ b_0^{-1}b_3b_4$&$ b_1b_2^{-1}b_3^{-1}$&$ b_0^{-1}b_2^{-1}b_4$&$ b_0^{-1}b_2^{-1}b_3^{-1}$\\&&&&&&\\
$s$& $\frac32$& $\frac12$& $\frac12$& $-\frac12$&  $-\frac12$& $-\frac32$\\&&&&&&\\
$\vec l$&$ (1,3,4)$&$ (1,\underline{2},4)$&$ (\underline{0},3,4)$&$ (1,\underline{2},\underline{3})$&$ (\underline{0},\underline{2},4)$&$(\underline{0},\underline{2},\underline{3})$\\
&&&&&&\\
$ \jh$&$(0,\tfrac12,1,\tfrac12)$&$(0,\tfrac12,1,\tfrac12)$&$(0,\tfrac12,1,\tfrac12)$&$(0,\tfrac12,1,\tfrac12)$&$(0,\tfrac12,1,\tfrac12)$&$(0,\tfrac12,1,\tfrac12)$
\\
&&&&&&\\$\pi_b^0$&$ b_1b_4b_5$&$ b_1b_3^{-1}b_5$&$ b_0^{-1}b_4b_5$&$ b_1b_3^{-1}b_4^{-1}$&$ b_0^{-1}b_3^{-1}b_5$&$ b_0^{-1}b_3^{-1}b_4^{-1}$\\
&&&&&&\\
\hline\hline

\end{tabular}
\end{center}
}
\end{table}

The vector $\vec h$ is next broken into segments separated by vertical bars such that two entries within a segment satisfy $h_l+h_{l+1}=k$ and adjacent entries belonging to different segments (i.e., separated by a vertical bar) satisfy  $h_l+h_{l+1}>k$.\footnote{A similar construction is presented in \cite{Idz}, where segments and their separators are called domains and domain walls respectively. As the following discussion will make clear, the elementary domain walls n \cite{Idz} are like the particles of the level-1 paths in our description.} For instance, with the above $\vec h$, we have
\begin{align}\label{h4.4b}
\vec h=(\underline{3}\,|\,1,2\,|\,\underline{2} \,|\,{2},1,2,1,2,\cdots).
\end{align}
Denote the first entry of each segment containing an odd number of elements, underlined above, as $\ell_0,\ell_1,\cdots$.  These values specify the positions of the peaks and valleys in the corresponding RSOS path $\vec p=(p_0,\cdots ,p_N)$ with entries $0\leq p_i\leq k$ and ending at $p_N=2j$.  
The heights of the sequence of peaks and valleys are: $k-\ell_0,\ell_1,k-\ell_2,\ell_3,\cdots$, with $k-\ell_0$ being   the initial point. Since the first segment is necessarily of odd length, $\ell_0=k$, so that the path initial point is $p_0=0$. Clearly, these data uniquely specifies the RSOS path. The one associated to the energy vector (\ref{h4.4b}) is thus: $\vec p=(0,1,2,1)$. Clearly, $\vec p=2\jh$, with $\jh$ given in Table \ref{tab1}. Modulo a trivial re-scaling, the RSOS paths are thus the same in the two constructions.

As a second example, consider the path $\ga\in\mkp{3}_0$ given in (\ref{exemplemky}), the corresponding 
$\vec h$ and  $\vec p$ are
\begin{align}
\vec h&= (\underline{3}\,|\,\underline{3}\,|\,\underline{2}\,|\, \underline{3}\,|\,\underline{1},2,1\,|\,\underline{3}\,|\,3,0,3,0\cdots)
\qquad \text{and}\qquad\vec p=(0,1,2,3,2,1,2,3,2,3,2,1,0).
\end{align}
The RSOS path $\vec p$ is precisely the one obtained from the operator construction of $\ga$ and pictured in the last figure of Fig. \ref{construction ci}, considered now with integer vertical units.

A third example is furnished by the path of Fig. \ref{exS4}, whose vector $\vec h$ is given below the figure. The corresponding sequence of peak and valley heights (including the initial and final points) is: $0,3,0,3,1,4,0,4,1,3,2$, leading to  the RSOS path of Fig. \ref{rsos4}, which is thus the same as $2\jh$.

These three examples suggest that the construction in \cite{Ara} and the one presented here associate the same RSOS path to a given $\mkp{k}$ path. The precise connexion between the two approaches will now be unraveled from the consideration of a fourth example.

Let us rework another example lifted from \cite{Ara}, still with $k=3$ and $j=\frac12$ (their example 3.1). We consider thus the following energy vector $\vec h$: 
\begin{align}\label{h3.1}
\vec h&= (\underline{3}\,|\,1,2,1,2\,|\,2,1\,|\,\underline{3},0,3\,|\,\underline{2},1,2\,|\,\underline{3},0,3\,|\,1,2\,|\,\underline{3},0,3\,|\,1,2,1,2,\cdots).
\end{align}
The RSOS path corresponding  to (\ref{h3.1}) is
\begin{align} \vec p=(0,1,2,3,2,1,2,3,2,1,0,1).
\end{align}
Let us now construct a path $\ga$ compatible with this energy vector. We consider the particular path constructed from right to left by choosing always the row of $H$ that is in lowest position, generating thus the path with maximal value of $n_-$. Keeping the vertical bars at the same positions as in $\vec h$, 
we have:
\begin{align}
\ga&=(-3\,|1,-1,1,-1\,|-1,1|-3,3,-3\,|-1,1,-1\,|-3,3,-3\,|1,-1\,|-3,3,-3\,|1,-1,\cdots).
\end{align}
The corresponding vector $\vec a$ is constructed by (\ref{gavsa}) and it is (still maintaining the vertical bars):
\begin{align}\label{a-}
\vec a&=(-1
\,|\,0,0,0,-1\,|\,0,-1\,|\,0,0,-2\,|\,0,0,-2\,|\,0,0,-1\,|\,0,-2\,|\,0,0,-1\,|\,0,0\cdots)
\end{align}
so that $n_-=11$ and $n_+=0$. 
The associated  vector $\vec l$ is 
\begin{align}
\vec l =  (\underline{0},\underline{4},\underline{6},\underline{9},\underline{9},\underline{12},\underline{12},\underline{15},\underline{17},\underline{17},\underline{20}),
\end{align}
so that  the corresponding $\jh$ reads
\begin{align} \jh=(0,\tfrac12,1,\tfrac32,1,\tfrac12,1,\tfrac32,1,\tfrac12,0,\tfrac12).
\end{align}
With each entry multiplied by 2, this matches the above RSOS path.

Now, the relation between the positions of the vertical bars and those where the $a_l\ne0$ in the vector (\ref{a-}) is manifest: all non-zero entries are those just before the segment separators. Moreover, all the values of $a_l$ for this particular path with mostly negative entries (i.e., with $n_-$ maximized) are precisely given by\footnote{For another illustration of this relationship (\ref{avsh}), consider the sixth $\vec a$ in Table \ref{tab1}.}
\begin{align}\label{avsh}
a_l=k-h_l-h_{l+1}.
\end{align}

The relation (\ref{avsh}) gives thus a unique and well defined procedure for constructing the vector $ \vec a$ associated to the operator
sequence, say $\pi_b^-$, with maximal $n_-$, given a fixed number of operators, out of which the vector $\jh$ is constructed. The position of the peaks and valleys for the $\jh$ path are identified, within the corresponding vector $\vec l$, by changes in the parity of successive underlined entries $l_i$: these correspond precisely to the positions of the $a_l\ne0$ preceding segments of odd length.
This leads to the same general procedure for constructing the RSOS path as described in \cite{Ara}. 

Now, as already said, to a given $\vec h$, there correspond many paths $\ga$, all with the same local energies.
 In \cite{Ara}, the values of the sector of these different paths are given by a precise combinatorial procedure. From our point of view, given $\vec h$, we construct the sequence $\pi^-_b$ of  maximal value of $n_-$ and all other sequences with same weight and same total number of operators are obtained recursively by proceeding as follows. In $\pi^-_b$ we replace the operators $b^{-1}_l$ by $b_{l+1}$, one by one.
  This operation is  weight preserving. Indeed the sector increases by 1, modifying $w$ by $-\frac12$, while the piece $\frac12\sum l|a_l|$ increases by $\frac12$. This operation is subject to one constraint: the total number of operators must be preserved at each step. This prevents annihilation processes such as $b_l^{-1}b^{-1}_{l+1}\to b_{l+1}b^{-1}_{l+1}=1$.
For instance, we generate the first five sequences $\pi_b$ of Table \ref{tab1} from the sixth one as (written here in terms of $\vec l$):
\begin{align}
(\underline{0},\underline{2},\underline{3})&\to(\underline{0},\underline{2},4)\to(\underline{0},3,4)\to(1,3,4)\nonumber\\
&\searrow (1,\underline{2},\underline{3})\to (1,\underline{2},4).
\end{align}

\end{appendix}
\vskip0.3cm
\noindent {\bf ACKNOWLEDGMENTS}

We thank T.A. Welsh for his critical reading of the manuscript,  suggesting substantial simplification in some proofs (and in particular, for proposing the simple equivalence proof (\ref{tre1})-(\ref{tre3})), and detailed explanations concerning the work \cite{Ara}. This work is supported  by NSERC.

\end{document}